\documentclass[journal]{IEEEtran}

\ifCLASSINFOpdf
\usepackage[pdftex]{graphicx}
\DeclareGraphicsExtensions{.pdf,.jpeg,.png}
\else
\usepackage[dvips]{graphicx}
\fi
\usepackage{subfigure}
\usepackage{epstopdf}
\usepackage[cmex10]{amsmath}
\usepackage{amssymb}
\usepackage{wasysym}
\usepackage{relsize}
\usepackage{algorithm}
\usepackage{algorithmic}
\usepackage{array}
\usepackage{cite}
\usepackage{color}
\usepackage{url}
\usepackage{bm}
\usepackage{enumerate}
\usepackage{epsfig,latexsym}
\usepackage{cite}
\usepackage{booktabs}
\usepackage{makecell} 

\begin{document}
\title{REMAA: Reconfigurable Pixel Antenna-based Electronic Movable-Antenna Arrays for Multiuser Communications}

\author{Kangjian~Chen,~\IEEEmembership{Student~Member,~IEEE},  Chenhao~Qi,~\IEEEmembership{Senior~Member,~IEEE},\\ Yujing Hong,~\IEEEmembership{Student~Member,~IEEE},  and Chau Yuen,~\IEEEmembership{Fellow,~IEEE}
	  \thanks{This work was supported in part by	the National Natural Science Foundation of China. Part of this work is to be submitted to the IEEE Global Communications Conference, Taipei, Taiwan, Dec. 2025~\cite{GlobeCom25CKJ3}. (\textit{Corresponding author: Chenhao~Qi})}
		\thanks{Kangjian~Chen and Chenhao~Qi are with the School of Information Science and Engineering, Southeast University, Nanjing 210096, China (e-mail: \{kjchen, qch\}@seu.edu.cn).}
		\thanks {Yujing Hong and Chau Yuen are with the School of Electrical and Electronics Engineering, Nanyang Technological University, Singapore 639798 (e-mail:yujing001@e.ntu.edu.sg; chau.yuen@ntu.edu.sg). 
		}
	
}

{}

\maketitle

\begin{abstract}
In this paper, we investigate reconfigurable pixel antenna (RPA)-based electronic movable antennas (REMAs) for multiuser communications. First, we model each REMA as an antenna characterized by a set of predefined and discrete selectable radiation positions within the radiating region. Considering the trade-off between performance and cost, we propose two types of REMA-based arrays: the partially-connected RPA-based electronic movable-antenna array (PC-REMAA) and fully-connected REMAA (FC-REMAA). Then, we formulate a multiuser sum-rate maximization problem subject to the power constraint and hardware constraints of the PC-REMAA or FC-REMAA. To solve this problem, we propose a two-step multiuser beamforming and antenna selection scheme. In the first step, we develop a two-loop joint beamforming and antenna selection (TL-JBAS) algorithm. In the second step, we apply the coordinate descent method to further enhance the solution of the TL-JBAS algorithm. In addition, we revisit mechanical movable antennas (MMAs) to establish a benchmark for evaluating  the performance of REMA-enabled multiuser communications, where MMAs can continuously adjust the positions within the transmission region. We also formulate a sum-rate maximization problem for MMA-enabled multiuser communications and propose an alternating beamforming and antenna position optimization scheme to solve it. Finally, we analyze the performance gap between REMAs and MMAs. Based on Fourier analysis, we derive the maximum power loss of REMAs compared to MMAs for any given position interval. Specifically, we show that the REMA incurs a maximum power loss of only 3.25\% compared to the MMA when the position interval is set to one-tenth of the wavelength. Simulation results demonstrate the effectiveness of the proposed methods.
\end{abstract}




\section{Introduction}
With the emergence of applications such as the Internet of Things, smart homes, and industrial automation, wireless communications face increasing performance demands, including higher data rates, broader coverage, lower power consumption, and greater adaptability~\cite{yang2019}. As a result, enhancing the capability, efficiency, and flexibility of wireless communication systems has become a primary focus of current research.

As the fundamental medium for information transmission, wireless channels play a crucial role in determining the performance of wireless communications~\cite{rappaport2001wireless}. Accordingly, optimizing channel utilization has emerged as a key research objective.  For example, conventional MIMO communications enhance transmission performance by exploiting the inherent degrees of freedom of channels~\cite{TWC05RW}.
Specifically, spatial multiplexing improves communication rates by enabling parallel transmission of multiple data streams while spatial diversity enhances communication reliability by exploiting the independent transmission of multiple channel paths. To further extend the spatial dimensions available for communication, massive MIMO systems have been developed by substantially increasing the number of antennas relative to conventional MIMO~\cite{JSTSP14LL,OBETrans,TWC20QCH}, which changes the channel characteristics and expands the flexibility available for signal design. In this regard, beamforming is performed to adapt to the  channel characteristics and exploit these additional flexibility~\cite{Ali2017Beamforming,JSTSP24CKJ,TVT24QCH}. By optimizing the utilization of channels, conventional MIMO and massive MIMO significantly enhance communication performance.

Notably, both the conventional and massive MIMO typically employ fixed-position antennas (FPAs), which limits their design flexibility. Specifically, the received signal power varies across the reception region due to the interference among signals from multiple paths. Consequently, the antenna positions can significantly impact communication performance, as they influence the strengths of received signal powers. In the case of FPAs, the received power of the antennas remains fixed under given channel conditions and cannot be enhanced through optimization, thereby limiting their communication performance. To address this issue, the antenna selection (AS) has been developed~\cite{CM04SS,TWC05MAF,TWC18Ass}. By deploying a large number of candidate antennas and selecting part of them for wireless communication, AS can adapt to instantaneous channel state information, thus overcoming the limitations of FPAs. Despite its advantages, AS presents two main challenges. On the one hand, to fully exploit the wireless channel, AS requires the deployment of a large number of antennas, which incurs significant hardware costs. Additionally, the physical size constraints of the antennas impose a minimum spacing between candidate antennas, limiting AS to  exploiting channels of specific discrete locations for wireless communications.

To address these challenges, recent research has explored novel methods for achieving flexible mobility of radiating elements, including the use of movable antennas (MAs), fluid antennas (FAs) and flexible intelligent metasurfaces (FIMs)~\cite{Zhu2024Historical,WCL24YSJ,WCL24FBQ,WCL24MWD,TWC25AJC,an2025flexible}. The key advantage of MAs/FAs/FIMs lies in their ability to move freely within a defined spatial region. Compared to FPAs, MAs/FAs/FIMs provide greater design flexibility by enabling the joint optimization of radiating element positions and beamforming vectors. Compared to AS, MAs/FAs/FIMs allow more flexible radiating element position adjustments, leading to greater improvements in communication performance.

Due to the significant potential of MAs/FAs/FIMs, substantial research efforts have been dedicated to exploring their characteristics~\cite{TWC24ZLP,TWC21WKK}. For example, in~\cite{TWC24ZLP}, the modeling and performance analysis of one MA are investigated, which reveals that the multipath channel gain exhibits periodic behavior in a given spatial region under deterministic channels. In~\cite{TWC21WKK}, the authors analyze the probability density function  and cumulative distribution function  of the signal envelopes for FAs under spatially correlated Rayleigh fading channels. Beyond these theoretical analyses, extensive research has also explored the practical potential of MAs/FAs/FIMs in wireless communication systems~\cite{TWC21WKK,TCom24XH,TWC24MWY,TWC24ZLP2}. In~\cite{TWC21WKK}, an approximate closed-form expression for the outage probability of FA systems is derived,  demonstrating that FAs can achieve arbitrarily small outage probabilities. In~\cite{TCom24XH}, the authors optimize both the transmit covariance matrices and the antenna position vectors of users to maximize system capacity, showing that FAs can significantly improve the capacity of multiple access channels. In~\cite{TWC24MWY}, a MIMO communication system with MAs is considered, indicating that MA systems substantially outperform  FPA systems in terms of MIMO channel capacity. In~\cite{TWC24ZLP2}, multiuser communication with MAs is investigated, demonstrating that the total transmit power can be significantly reduced compared to conventional FPA systems.  In addition to these advancements, the effectiveness of MAs and FAs has also been validated in other systems, including wireless sensing~\cite{TWC24MWY2}, integrated sensing and communications~\cite{Ma2025Movable,TWC25}, secure wireless communications~\cite{SPL24HGJ}, and symbiotic radio systems~\cite{WCL24ZC}.

Although extensive research has demonstrated the promising potential of MAs/FAs/FIMs, their practical implementation  remains an open issue. Initially, MAs/FAs/FIMs are controlled through mechanical methods, where FAs rely on liquid pumping or pressure regulation to adjust the antenna  positions, MAs use  mechanical devices such as motors to reposition antenna elements within the transmission region, and FIMs use micro-mechanical mechanisms to reconfigure the positions of the radiating elements. While these methods enable flexible movement of radiating elements, their dependence on mechanical adjustments results in slow position updates,  which renders them unsuitable for rapidly varying channels. Recently, an implementation of MAs/FAs based on reconfigurable pixel antennas (RPAs) is proposed in \cite{OJAP25ZJ}.  By electronically controlling radio frequency (RF) components between pixels, this method enables rapid changing of antenna positions,  and offers a feasible solution for the practical realization of MAs/FAs.

In this paper, we investigate RPA-based electronic movable antennas (REMAs), as proposed in \cite{OJAP25ZJ}, for multiuser sum-rate maximization through joint beamforming and antenna selection. The main contributions of this paper are summarized as follows.
\begin{itemize}
\item First, we model each REMA as an antenna characterized by a set of predefined and discrete selectable radiation positions within the radiating region. Considering the trade-off between performance and cost, we propose two types of REMA-based arrays: the partially-connected RPA-based electronic movable-antenna array (PC-REMAA) and fully-connected RPA-based electronic movable-antenna array (FC-REMAA).

\item Then, we investigate multiuser communications, where the base station (BS) is equipped with either a PC-REMAA or an FC-REMAA, while users employ FPAs. We formulate a multiuser sum-rate maximization problem subject to power constraint and hardware constraints of the PC-REMAA or FC-REMAA. To solve this problem, we propose a two-step multiuser beamforming and antenna selection (TS-MBAS) scheme. In the first step, we develop a two-loop joint beamforming and antenna selection (TL-JBAS) algorithm. In the second step, we apply the coordinate descent method to further enhance the solution of the TL-JBAS algorithm.

\item In addition, we revisit mechanical movable antennas (MMAs) to establish a benchmark for evaluating  the performance of REMA-enabled multiuser communications, where MMAs can continuously adjust the positions within the transmission region.  We also formulate a sum-rate maximization problem for MMA-enabled multiuser communications and propose an alternating beamforming and antenna position optimization (ABAPO) scheme to solve it.

\item Moreover, we analyze the performance gap between REMAs and MMAs. Specifically, we transform the received signal of MMAs into the discrete-time Fourier transform (DTFT) of the channel coefficients. Based on Fourier analysis, we derive the maximum power loss of REMAs compared to MMAs for any given position interval.

\end{itemize}
\begin{figure}[t]
	\centering
	\includegraphics[width=80mm]{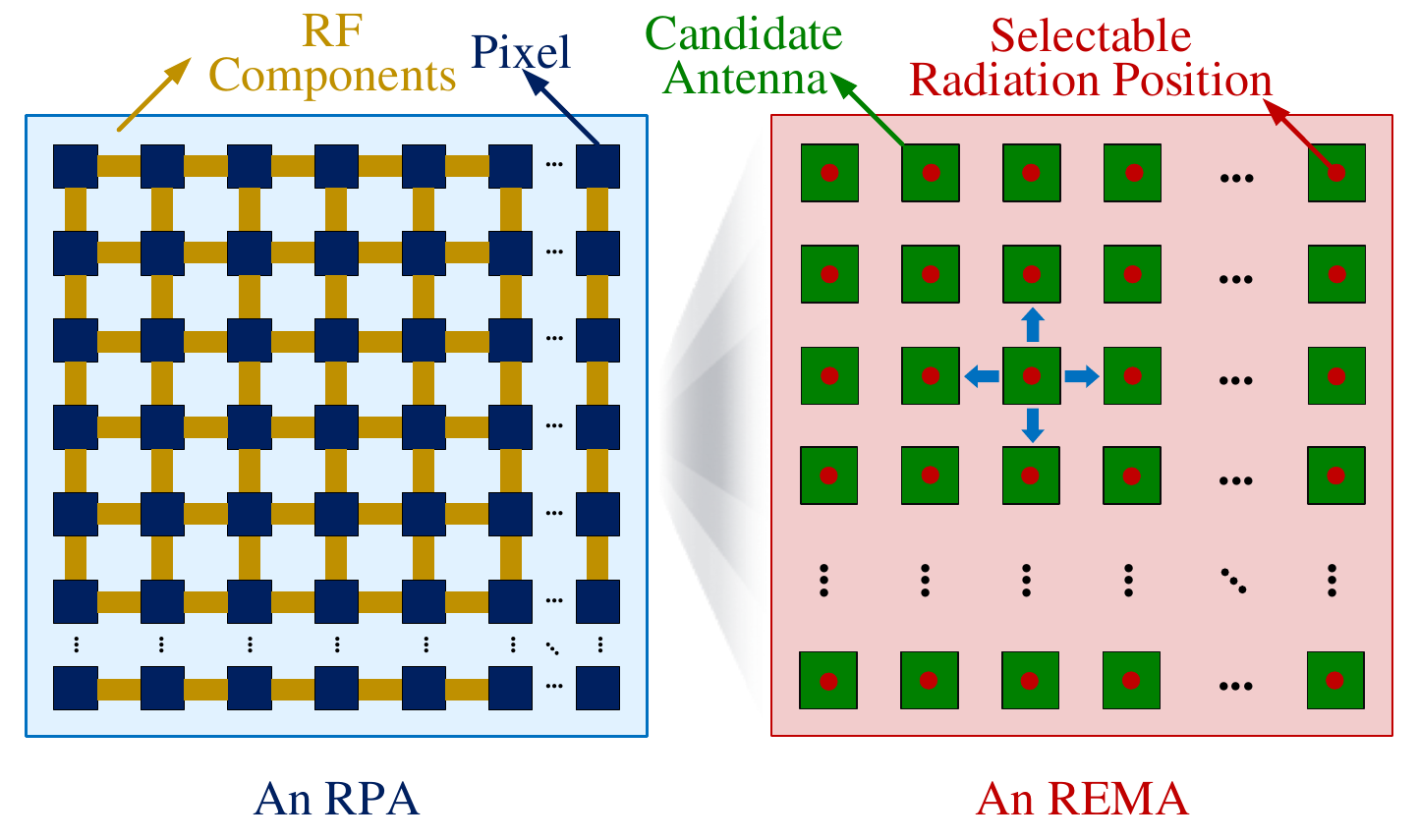}
	\caption{Illustration of an RPA and an RMAA.}
	\label{fig_image1}
	\vspace{-0.5cm}
\end{figure}

The rest of this paper is organized as follows. The electronic movable antennas are introduced in Section~\ref{ElectronicMovableAntennas}. The system model and problem formulation are presented in Section~\ref{SystemModel}.  The multiuser beamforming  and antenna selection for REMAs  is developed in Section~\ref{EMA}. The revisit of the MMAs is given in Section~\ref{MMA}. The performance gap analysis between REMAs and MMAs is discussed in Section~\ref{Analysis2}.  The proposed methods are evaluated in Section~\ref{SR}, and the paper is concluded in Section~\ref{CC}.

\textit{\textbf{Notations:}} Lowercase and uppercase bold symbols denote vectors and matrices, respectively. $[\boldsymbol{A}]_{m,:}$,  $[\boldsymbol{A}]_{:,n}$, and $[\boldsymbol{A}]_{m,n}$ denote the $m$th row, the $n$th column, and the entry on the $m$th row and the $n$th column of a matrix $\boldsymbol{A}$. $\boldsymbol{a} \odot\boldsymbol{b}$ denotes the Hadamard  product of vectors $\boldsymbol{a}$ and $\boldsymbol{b}$. $[\boldsymbol{a}]_m$ denotes the $m$th entry of a vector $\boldsymbol{a}$. $|\cdot|$, $\|\cdot\|_2$, and $\|\cdot\|_{\rm F}$ denote the absolute value of a scalar, the $\ell_2$-norm of a vector,  and the Frobenius norm  of a matrix, respectively. $(\cdot)^{\rm T}$, $(\cdot)^{\rm H}$, and $(\cdot)^{\dagger}$ denote the transpose, Hermitian transpose, and the Moore-Penrose inverse, respectively.  $\mathbb{C}$  and $\mathcal{CN}(\cdot)$ denote the set of complex numbers and complex Gaussian distribution, respectively. $\mathrm{vec}\{\boldsymbol{A}\}$ denotes the vectorization of the matrix $\boldsymbol{A}$. $\mathcal{R}\{a\}$ denotes the real part of a complex number $a$. $\otimes$ denotes the Kronecker product.

\section{Electronic Movable Antennas}\label{ElectronicMovableAntennas}
In this section, we first introduce the implementation of REMAs. Then, we discuss the design of antenna arrays using REMAs and propose the PC-REMAA and FC-REMAA.
\subsection{RPA-based Electronic Movable Antennas}\label{EMARPA}
As shown on the left of Fig.~\ref{fig_image1}, in RPAs, the radiating surface is composed of  multiple pixel elements, where  RF components, such as PIN diodes, are employed between these pixel elements to adjust their connections. As demonstrated in \cite{TAP14_SSC,TAP22_JLW}, by dynamically controlling the states of these  RF switches, the current distribution across the radiating surface can be altered and thus the radiation pattern can be modified.  It is further established in \cite{OJAP25ZJ} that changing radiation patterns through this method is equivalent to physically moving antenna positions. As a result, RPAs can effectively realize movable antennas by electronically controlling the RF switches. We refer to antennas implemented in this manner as REMAs.

Based on the implementation of RPAs, we model each REMA as an antenna characterized by a set of predefined and discrete selectable radiation positions within the radiating region, as shown on the right of Fig.~\ref{fig_image1}. Specifically, when an REMA selects a discrete radiation position, it radiates from that position, which is then regarded as a candidate antenna. The parameters of each REMA, including the sizes, positions, numbers, and spacings of candidate antennas, are adjustable during manufacturing, which can be achieved by designing the RPAs with the dedicated efforts of antenna researchers. Once manufactured and deployed, these parameters become fixed and cannot be easily changed. In this work, the candidate antennas are arranged in a uniform planar configuration for simplicity. However, this approach can also be extended to other practical candidate antenna arrangements in future designs.

\textit{\textbf{Remark 1:}}  Unlike conventional AS, which relies on selecting from a predefined set of physically separated antennas with fixed spacing (e.g., half wavelength or larger), the REMA achieves finer position adjustment through collaborative reconfiguration of multiple pixel elements. Specifically, the RPA-based REMA allows candidate antennas to be densely arranged (e.g., 12 ports within half wavelength in~\cite{OJAP25ZJ}), enabling sub-wavelength-level spatial resolution. This is fundamentally distinct from AS, as the REMA dynamically synthesizes radiation patterns by controlling interconnected pixel states rather than merely selecting from isolated antennas. Consequently, the REMAs offer enhanced flexibility in exploiting spatial channel variations while avoiding the hardware complexity and physical size limitations of traditional AS.

\subsection{RPA-based Electronic Movable-Antenna Arrays}
In this part, we discuss the  design of electronic movable-antenna arrays with the REMAs. 

\begin{figure}[!t]
	\centering
	\includegraphics[width=80mm]{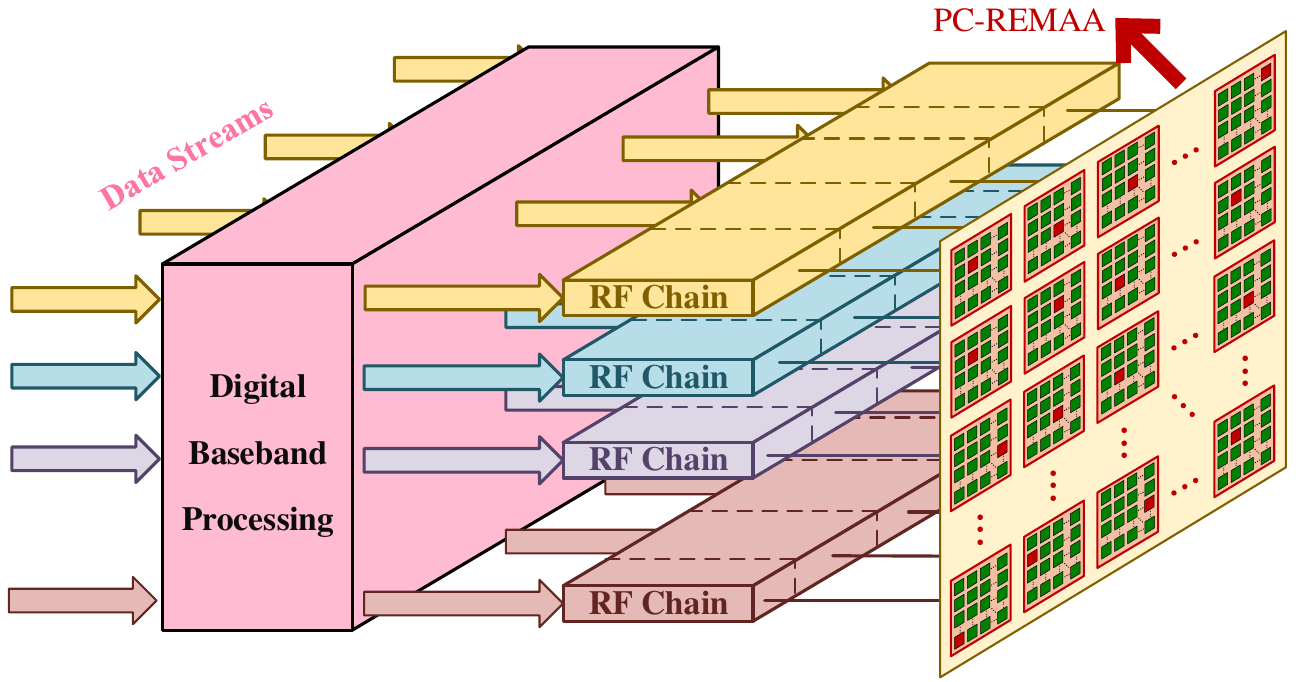}
	\caption{Illustration of the signal processing architecture for the PC-REMAA.}
	\label{fig_image2}
\end{figure}

\begin{figure}[!t]
	\centering
	\includegraphics[width=81.5mm]{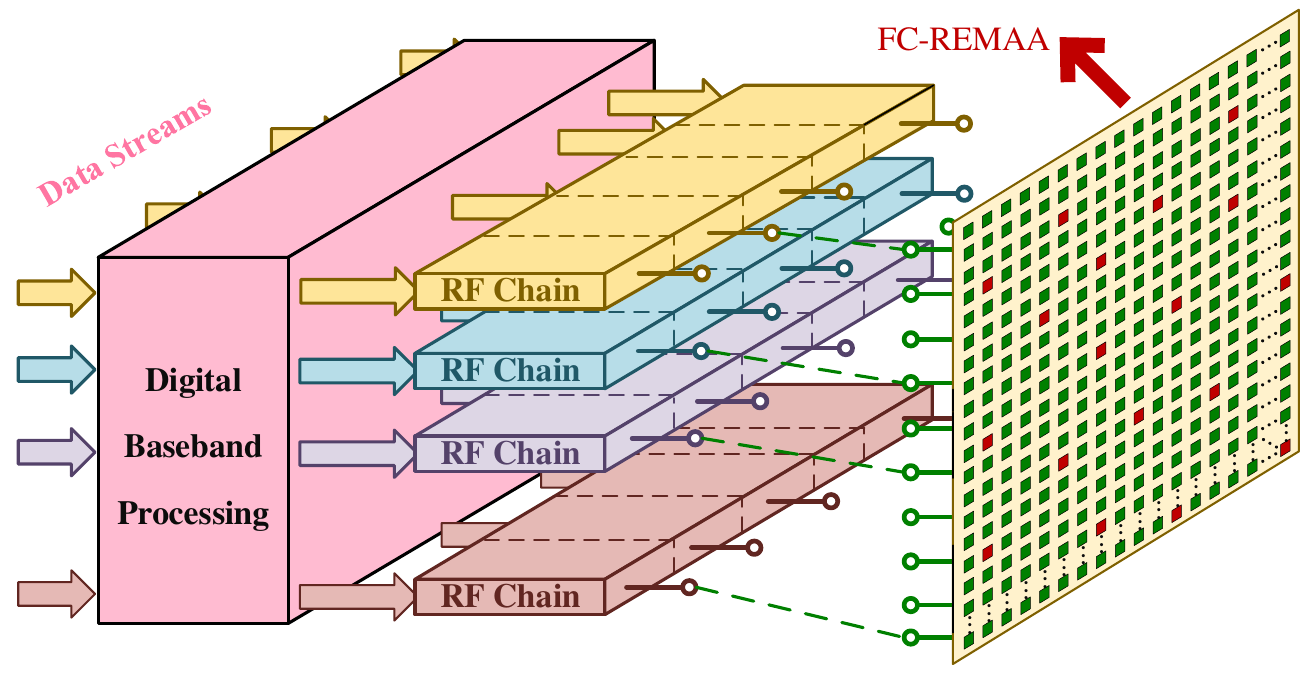}
	\caption{Illustration of the signal processing architecture for  the FC-REMAA.}
	\label{fig_image3}
		\vspace{-0.5cm}
\end{figure}

A  straightforward method for constructing a movable-antenna array is to combine multiple REMAs together, as depicted in Fig.~\ref{fig_image2}. To achieve a simplified and cost-effective RF architecture, each REMA is solely connected to a single RF chain.  This configuration is referred to as the PC-REMAA. Due to the connection constraints between the RF chains and antennas, only one candidate antenna is activated at a time for each REMA in PC-REMAA. In other words, each RF chain can only select one antenna from the set of candidate antennas in the REMA it is connected to.

To enhance the design flexibility of the PC-REMAA, a configuration called the FC-REMAA is further proposed, as illustrated in Fig.~\ref{fig_image3}. In this architecture, the entire antenna array is implemented using an RPA. By dynamically controlling the states of RF switches between pixels, multiple candidate antennas are selected at a time and then connected to the RF chains. In the FC-REMAA, each RF chain has the capability to connect to any of the selected candidate antennas. To facilitate this flexibility, an additional switch network is required to manage the connections between the RF chains and the dynamically selected antennas.

\textit{\textbf{Remark 2: }} The comparison between PC-REMAA and FC-REMAA demonstrates distinct trade-offs in hardware complexity, power efficiency, and system performance. PC-REMAA offers simpler deployment and lower power consumption but is limited in flexibility and performance. In contrast, FC-REMAA provides better performance and greater flexibility at the cost of higher design complexity and power consumption. Specifically, FC-REMAA requires simultaneous activation of multiple candidate antennas, which demands careful antenna design to manage mutual coupling and maintain impedance matching. The addition of a switch network to enable flexible connections between RF chains and candidate antennas further increases complexity, insertion loss, and power consumption. Additionally, the scalability of FC-REMAA is challenged by the quadratic growth of switch network complexity as the number of antennas and RF chains increases. PC-REMAA simplifies hardware implementation by connecting one REMA with an RF chain. The fixed one-to-one mapping reduces circuit complexity, insertion loss, and power consumption. However, this limits flexibility, as each RF chain can only access one candidate antenna per REMA, reducing spatial diversity and beamforming capabilities compared to FC-REMAA, particularly in complex beamforming or rapid spatial adaptation scenarios. These trade-offs suggest that PC-REMAA is suitable for cost-sensitive deployments with moderate performance needs, while FC-REMAA is better suited for high-performance systems requiring flexibility and capacity optimization.

\begin{figure}[!t]
	\centering
	\includegraphics[width=75mm]{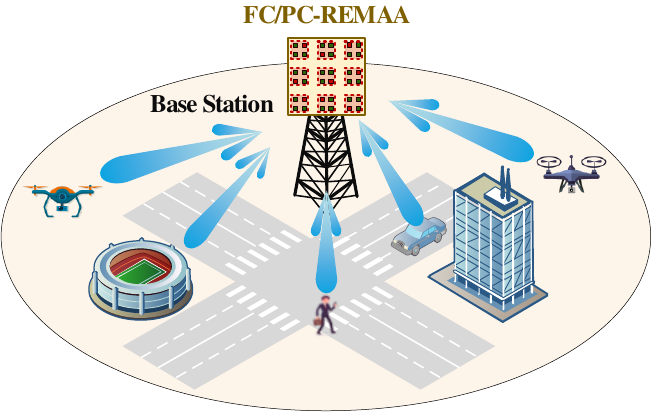}
	\caption{Illustration of the multiuser communication system with REMAAs.}
	\label{Scenarios}
\end{figure}

\section{System Model and Problem Formulation}\label{SystemModel}
\subsection{System Model}
As illustrated in Fig.~\ref{Scenarios}, we consider a multiuser communication system, where a BS employs a PC-REMAA or an FC-REMAA to serve $K$ single-antenna users. Without loss of generality, we assume the antennas of both the PC-REMAA and FC-REMAA are arranged in a uniform planar array. If the PC-REMAA is employed, the transmit array includes $M_{\rm r}$ rows and  $M_{\rm c}$ columns of REMAs. From Section~\ref{EMARPA}, by controlling the connections between pixels, each REMA can adjust the radiation pattern to change the radiation position for signal transmission and thus form a series of candidate antennas. Suppose each REMA can support $S_{\rm r}$ rows and $S_{\rm c}$ columns of candidate antennas. As a result, The total candidate antennas in each row and column of the PC-REMAA are $N_{\rm c} = M_{\rm c}S_{\rm c}$ and $N_{\rm r} = M_{\rm r}S_{\rm r}$, respectively. Then, the total number of candidate antennas in the PC-REMAA is  $N_{\rm t} = N_{\rm r}N_{\rm c}$.  On the other hand, if the FC-REMAA is employed, the whole antenna array can be taken as a large-scale RPA. We assume candidate antennas in each row and each column of the FC-REMAA are the same as those of the PC-REMAA for simplicity.  Then, the whole FC-REMAA also includes $N_{\rm t} = N_{\rm r}N_{\rm c}$ candidate antennas in total.  The spacing between adjacent candidate antennas for both the PC-REMAA and the FC-REMAA is denoted as $d_{\rm c}$ while the spacing between adjacent REMAs for the PC-REMAA is denoted as $d_{\rm e}$. In this work, we focus on the design of the REMAs at the BS and the positions of antennas at the users are assumed to be fixed.

\begin{figure}[!t]
	\centering
	\includegraphics[width=65mm]{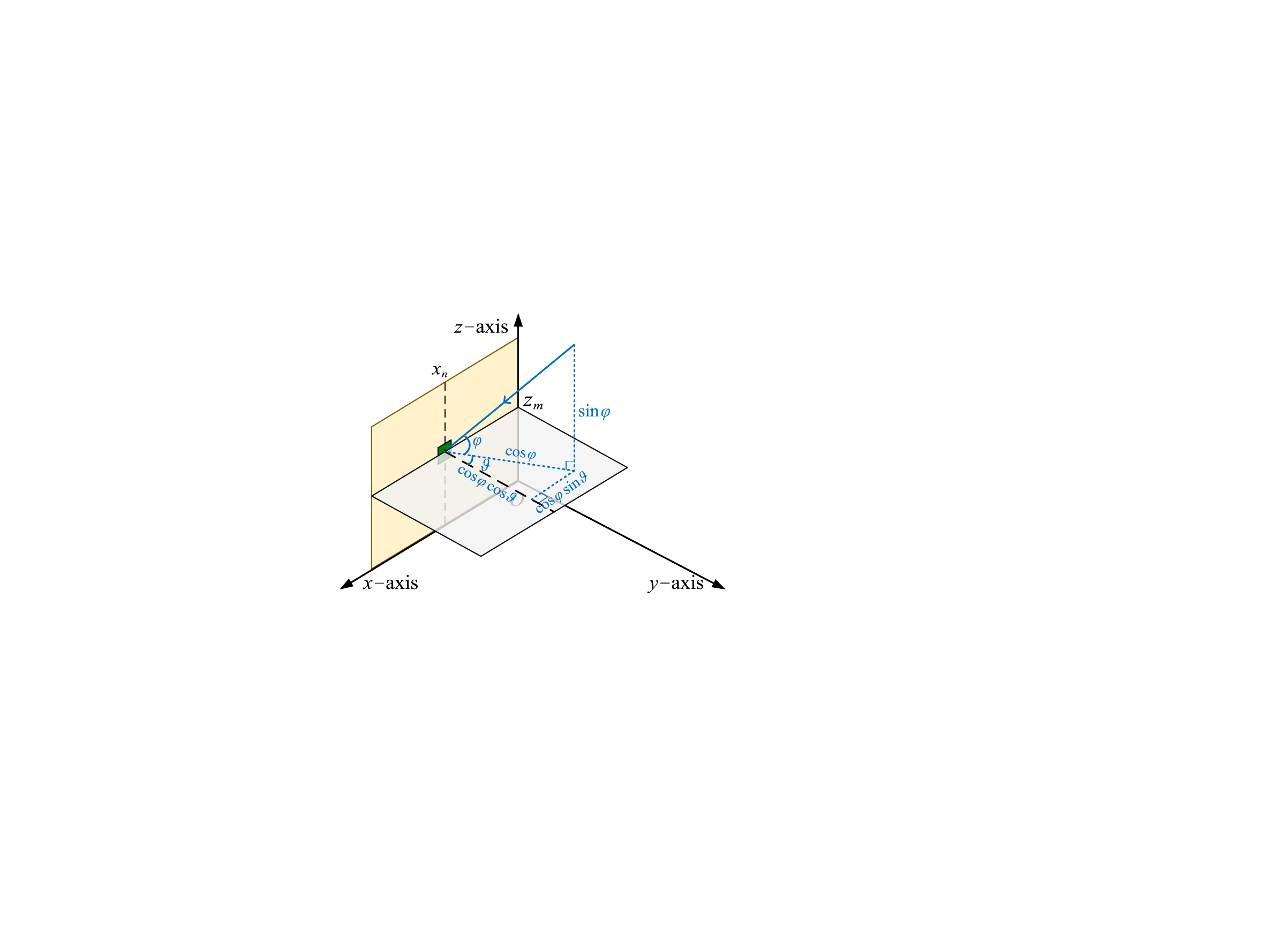}
	\caption{Illustration of the spatial angles in Cartesian coordinate system.}
	\label{fig_image5}
\end{figure}

For  REMA-based multiuser communications, the channels between the BS and the users are determined by propagation environments and the positions of candidate antennas. To represent  the positions of candidate antennas, in Fig.~\ref{fig_image5}, we establish a Cartesian coordinate system with the antenna located at the $N_{\rm r}$th row and the $N_{\rm c}$th column as the origin. Denote the coordinate of the candidate antenna located at  the $m$th row and the $n$th column as $[x_n,0,z_m]$. The steering vector for the channel path with azimuth angle $\vartheta$ and elevation angle $\varphi$ can be expressed as 
\begin{align}
	\boldsymbol{\alpha}(N_{\rm t},\vartheta,\varphi) = \sqrt{\frac{1}{N_{\rm t}}} e^{j2\pi \boldsymbol{x}\cos \varphi \sin \vartheta/\lambda}\otimes e^{j2\pi \boldsymbol{z}\sin \varphi/\lambda},
\end{align}
where $\lambda$ denotes the carrier wavelength, $\boldsymbol{x}\triangleq[x_1,x_2,\cdots,x_{N_{\rm c}}]^{\rm T}$, and $\boldsymbol{z}\triangleq[z_1,z_2,\cdots,z_{N_{\rm r}}]^{\rm T}$. Then, the channel between the BS and the $k$th user can be expressed~as 
\begin{align}\label{channel}
	\boldsymbol{h}_{k} = \sum_{l=1}^{L_k}\gamma_{k}^{(l)}\boldsymbol{\alpha}\big(N_{\rm t},\theta_k^{(l)},\phi_k^{(l)}\big),
\end{align}  
where $L_{k}$ denotes the number of channel paths, $\gamma_{k}^{(l)}$, $\theta_k^{(l)}$ and $\phi_k^{(l)}$ denote the channel gain, the azimuth angle and the elevation angle of the $l$th path, respectively. For the PC-REMAA, $x_n$ and $z_m$ can be expressed as 
\begin{align}\label{PREMAPosition}
	x_n = (M_{\rm c}-q)d_{\rm e} + (S_{\rm c}-t)d_{\rm c},\nonumber\\
	z_m = (M_{\rm r} -p)d_{\rm e} + (S_{\rm r}-s)d_{\rm c},
\end{align}
where we assume that the antenna, located at the $s$th row and $t$th column of the REMA in the $p$th row and $q$th column, corresponds to the antenna at the $m$th row and $n$th column of the PC-REMAA, i.e., $m = (p-1)S_{\rm r} + s$ and $n = (q-1)S_{\rm c} + t$. According to the PC-REMAA settings, we have $p\in\{1,2,\cdots,M_{\rm r}\}$, $q\in\{1,2,\cdots,M_{\rm c}\}$, $s\in\{1,2,\cdots,S_{\rm r}\}$, and $t\in\{1,2,\cdots,S_{\rm c}\}$. For the FC-REMAA, $x_n$ and $z_m$ can be expressed as 
\begin{align}\label{FREMAPosition}
	x_n = (N_{\rm c} - n)d_{\rm c},~\mbox{and}~z_m = (N_{\rm r} - m)d_{\rm c}.
\end{align}
Substituting $x_n$ and $z_m$ in \eqref{PREMAPosition} and \eqref{FREMAPosition} into \eqref{channel}, we  can obtain the channels for the PC-REMAA and FC-REMAA, respectively.

\subsection{Problem Formulation}
Due to the interference among signals from different channel paths, both the signal strengths and phases vary across the candidate antennas of the PC/FC-REMAA. Usually, greater signal strength corresponds to a higher received signal-to-noise ratio (SNR) and thus leads to better communication quality, which indicates that selecting antennas with stronger signals results in better communication performance.  In addition, due to variations in channel state information, the optimal antenna configuration may also change over time. Thus, dynamic antenna selection is required to adapt to these varying channel conditions and optimize communication performance. Denote the number of selected antennas and RF chains as $N_{\rm s}$ and $N_{\rm RF}$, respectively. To perform baseband processing, each of the selected antennas is solely connected to an RF chain. Then, we have $N_{\rm s}\le N_{\rm RF}$, limited by the number of RF chains. In addition, the number of selected antennas must exceed the number of users to support independent transmission, i.e., $N_{\rm s}\ge K$.

Denote the selection matrix as $\boldsymbol{T}\in\mathbb{Z}^{N_{\rm r}\times N_{\rm c}}$, where each entry is one or zero, i.e.,
\begin{align}\label{binary}
	[\boldsymbol{T}]_{m,n} \in \{0,1\}.
\end{align} 
If $[\boldsymbol{T}]_{m,n}$ is one, the candidate antenna located at the $m$th row and $n$th column is selected for data transmission; otherwise, it is not selected. Since $N_{\rm s}$ antennas are selected, we have
\begin{align}\label{sum3}
	\sum_{m=1}^{N_{\rm r}}\sum_{n=1}^{N_{\rm c}} [\boldsymbol{T}]_{m,n} = N_{\rm s}.
\end{align}
To avoid coupling effects between selected antennas, the minimum row and column spacing between selected antennas usually exceed $\lambda/2$, which can be easily converted to the minimum antenna index spacing $D$ by considering the configurations of PC/FC-REMAA. For example, we have $D = \lceil{\lambda/(2d_{\rm c})\rceil}$ antennas for the FC-REMAA. If an antenna located at the $m$th row and $n$th column is selected, then antennas within a square region centered at the selected antenna extending $D$ rows and columns in both directions, cannot be selected.  This restriction can be mathematically formulated as 
\begin{align}\label{sum1}
	\sum_{a = m-D}^{m+D}\sum_{b = n-D}^{n+D} [\boldsymbol{T}]_{a,b} \le1,
\end{align}  
for $m = D + 1,\cdots,N_{\rm r}-D$ and $n=D+1,\cdots,N_{\rm c}-D$. The constraint in \eqref{sum1} can be equivalently expressed in matrix form as
\begin{align}\label{sum1eq}
	\sum_{a=1}^{N_{\rm r}} \sum_{b=1}^{N_{\rm c}} [\boldsymbol{B}_{m,n}]_{a,b}[\boldsymbol{T}]_{a,b} \le 1,
\end{align}
where $\boldsymbol{B}_{m,n}\in\mathbb{Z}^{N_{\rm r}\times N_{\rm c}}$ is defined as
\begin{align}
	[\boldsymbol{B}_{m,n}]_{a,b} = \begin{cases}
		1,  a\!\in\![m\!-\!D,\!m\!+\!D],~\mathrm{and}~b\!\in\![n\!-\!D,\!n\!+\!D],\\
		0,~\mathrm{others}.\\
	\end{cases}
\end{align}
Note that $\boldsymbol{B}_{m,n}$ characterizes the selection constraints by marking positions within the exclusion zone with ones, while positions outside this region remain zero.

For the PC-REMAA, only one antenna can be selected from the candidate antennas within each REMA, as these candidate antennas are exclusively connected to a single RF chain. This constraint can be expressed as
\begin{align}\label{sum2}
	\sum_{m = (p-1)  S_{\rm r} + 1}^{pS_{\rm r}}\sum_{n = (q-1)S_{\rm c} +1}^{qS_{\rm c}} [\boldsymbol{T}]_{a,b} = 1,
\end{align}  
for $p = 1,\cdots,M_{\rm r}$ and $q = 1,\cdots,M_{\rm c}$. Similar to \eqref{sum1eq}, \eqref{sum2} can also be equivalently rewritten as 
\begin{align}\label{sum2eq}
	\sum_{a = 1}^{N_{\rm r}}\sum_{b = 1}^{N_{\rm c}}  [\boldsymbol{C}_{p,q}]_{a,b}[\boldsymbol{T}]_{a,b} = 1,
\end{align}  
where $\boldsymbol{C}_{p,q}\in\mathbb{Z}^{N_{\rm r}\times N_{\rm c}}$ is defined as
\begin{align}
	[\boldsymbol{C}_{p,q}]_{a,b} = \begin{cases}
		1,  \begin{array}{c}
			a\!\in\![(p-1)S_{\rm r} + 1,pS_{\rm r}]\\
			b\!\in\![(q-1)S_{\rm c} +1,qS_{\rm c}],\\
		\end{array}\\
		0,~\mathrm{others}.\\
	\end{cases}
\end{align}
Here, $\boldsymbol{C}_{p,q}$ characterizes the selection constraints for the PC-REMAA by marking the positions of candidate antennas within the REMA located at the $p$th row and $q$th column with ones, while all other positions remain zero.

To facilitate following formulations, we vectorize the selection matrix $\boldsymbol{T}$ as $\boldsymbol{t} \triangleq \mathrm{vec}\{\boldsymbol{T}\}$. Then the constraints in \eqref{binary}, \eqref{sum3}, \eqref{sum1}, and \eqref{sum2} can be respectively expressed as 
\begin{subequations}\label{Cons}
\begin{align}
&[\boldsymbol{t}]_p\in \{0,1\},~\mathrm{for}~p=1,2,\cdots,N_{\rm t}, \label{Cons1}\\
&\boldsymbol{1}_{N_{\rm t}}^{\rm T}\boldsymbol{t} = N_{\rm s}, \label{Cons2}\\
&\boldsymbol{S}^{\rm T}\boldsymbol{t}\le \boldsymbol{1}_{N_{\rm t}}, \label{Cons3}\\
&\boldsymbol{Q}^{\rm T}\boldsymbol{t}= \boldsymbol{1}_{M_{\rm t}}. \label{Cons4}
\end{align}
\end{subequations}
In \eqref{Cons3}, $\boldsymbol{S}\in\mathbb{Z}^{N_{\rm t}\times N_{\rm t}}$ denotes the spacing constraint matrix and can be expressed as 
\begin{align}
[\boldsymbol{S}]_{:,s} = \mathrm{vec}\{\boldsymbol{B}_{m,n}\},
\end{align}
for $s = (n-1)N_{\rm r} + m$, $m= 1,\cdots,N_{\rm r}$, and $n= 1,\cdots,N_{\rm c}$. In \eqref{Cons4}, $M_{\rm t}\triangleq M_{\rm r}M_{\rm c}$ denotes the number of REMAs in the PC-REMAA. $\boldsymbol{Q}\in\mathbb{Z}^{N_{\rm t}\times M_{\rm t}}$ denotes the partially-connected antenna selection constraint matrix and can be expressed as 
\begin{align}
	[\boldsymbol{Q}]_{:,s} = \mathrm{vec}\{\boldsymbol{C}_{p,q}\},
\end{align}
for $s = (q-1)M_{\rm r} + p$, $p= 1,\cdots,M_{\rm r}$, and $q= 1,\cdots,M_{\rm c}$. Then, we respectively denote the sets of $\boldsymbol{t}$ for the PC-REMAA and FC-REMAA as 
\begin{align}\label{ASconstaint}
	&\boldsymbol{\Phi}_{\rm P} = \{\boldsymbol{t}\in\mathbb{Z}^{N_{\rm t}}|\eqref{Cons1},~\eqref{Cons2},~\eqref{Cons3}, \mbox{and}~\eqref{Cons4}\},\nonumber \\
	&\boldsymbol{\Phi}_{\rm F} = \{\boldsymbol{t}\in\mathbb{Z}^{N_{\rm t}}|\eqref{Cons1},~\eqref{Cons2}, \mbox{and}~\eqref{Cons3}\}.
\end{align}

In this work, we aim at maximizing the sum-rate of \( K \) users by selecting \( N_{\rm s} \) antennas from the \( N_{\rm t} \) candidate antennas and optimizing the multiuser beamforming, subject to the constraints in \eqref{binary}, \eqref{sum3}, \eqref{sum1}, and \eqref{sum2}. The optimization problem can be formulated as  
\begin{subequations}\label{ProblemFormulation}
	\begin{align}
		&\max_{\boldsymbol{F},\boldsymbol{t}}~\sum_{k=1}^{K} R_k\label{PObj}\\
		&~\mathrm{s.t.}~~\left\|\boldsymbol{F}\right\|_{\rm F}^2 \le P,\label{Pb}\\
		&~~~~~~~~\boldsymbol{t}\in\boldsymbol{\Phi}_{\rm P}~\text{or}~\boldsymbol{\Phi}_{\rm F}.\label{Pcc} 
	\end{align}
\end{subequations}
In \eqref{PObj}, \( R_k \) represents the achievable data rate of the \( k \)th user and is given by  
\begin{align}
	R_k = \log_2\left(1+\frac{\big|\boldsymbol{h}_{k}^{\rm H}\boldsymbol{\overline{T}}\boldsymbol{f}_{k}\big|^2}{\sigma^2+\sum_{i =1,i\neq k}^{K}\big|\boldsymbol{h}_{k}^{\rm H}\boldsymbol{\overline{T}}\boldsymbol{f}_{i}\big|^2} \right),
\end{align}
where \( \boldsymbol{\overline{T}} \triangleq \mathrm{diag}\{\boldsymbol{t}\} \), \( \boldsymbol{f}_{k} \triangleq [\boldsymbol{F}]_{:,k} \), and \( \boldsymbol{F} \in \mathbb{C}^{N_{\rm t} \times K} \) denotes the multiuser transmit beamforming matrix.  Constraint in~\eqref{Pb} ensures that the total transmit power satisfies the power budget, i.e.,  $\left\|\boldsymbol{F}\right\|_{\rm F}^2 \le P$, where $P$ represents the maximum allowable transmit power.  In  \eqref{Pcc}, \( \boldsymbol{t} \) belongs to different feasible sets depending on the employed REMAA architecture: \( \boldsymbol{t} \in \boldsymbol{\Phi}_{\rm P} \) for PC-REMAA and \( \boldsymbol{t} \in \boldsymbol{\Phi}_{\rm F} \) for FC-REMAA.  

\section{Multiuser Beamforming  and Antenna Selection for REMAAs}\label{EMA}
In this section, we focus on solving \eqref{ProblemFormulation} for the design of multiuser beamforming  and antenna selection in PC/FC-REMAA, where a TS-MBAS scheme is proposed. In the first step of the TS-MBAS scheme, we propose a TL-JBAS algorithm, as elaborated from Section~\ref{PF} to Section~\ref{SCTL}. In the second step of the TS-MBAS scheme, we apply the coordinate descent method to further enhance the solution of the TL-JBAS algorithm, as elaborated in Section~\ref{enhancement}.
\subsection{Problem Conversion}\label{PF}
Note that \eqref{ProblemFormulation} involves a nonconvex objective function, as shown in \eqref{PObj}, and binary constraints, as defined in \eqref{binary}, which make the problem  a mixed-integer nonlinear programming (MINLP) problem. According to \cite{Lee2011},  MINLP problems are widely recognized as nondeterministic polynomial-time hard and it is  unlikely to achieve optimality in polynomial time. Therefore, we turn to convert \eqref{ProblemFormulation} into a tractable form to efficiently find a suboptimal solution.

First, we focus on the power constraint in \eqref{Pb}. Based on \textit{Proposition 3} in \cite{TSP23ZXT}, any nontrivial stationary point of the beamforming matrix 
$\boldsymbol{F}$ that maximizes the sum-rate will satisfy the power constraint with equality. This property allows us to simplify the problem by removing the power constraint from \eqref{Pb}. As a result, we can reformulate \eqref{ProblemFormulation} into a more tractable and  equivalent form as
\begin{subequations}\label{ProblemFormulation2}
	\begin{align}
		&\max_{\boldsymbol{F},\boldsymbol{t}}~\sum_{k=1}^{K} \widetilde{R}_k\label{PObj2}\\
		&~\mathrm{s.t.}~~\boldsymbol{t}\in\boldsymbol{\Phi}_{\rm P}~\mathrm{or}~\boldsymbol{\Phi}_{\rm F}, 
	\end{align}
\end{subequations}
where $\widetilde{R}_k$ can  be expressed as 
\begin{align}
	\widetilde{R}_k = \log_2\left(1+\frac{\big|\boldsymbol{h}_{k}^{\rm H}\boldsymbol{\overline{T}}\boldsymbol{f}_{k}\big|^2}{\frac{\sigma^2\|\boldsymbol{F}\|_{\rm F}^2}{P}+\sum_{i =1,i\neq k}^{K}|\boldsymbol{h}_{k}^{\rm H}\boldsymbol{\overline{T}}\boldsymbol{f}_{i}\big|^2} \right).
\end{align}

Then, we turn our attention to the nonconvex objective in \eqref{PObj2}. By applying \textit{Lemma 4.1} from \cite{TWC15SQJ}, we can address the objective function in \eqref{PObj2} and convert \eqref{ProblemFormulation2} into an equivalent form~as
\begin{subequations}\label{ProblemFormulation3}
	\begin{align}
		&\min_{u_k,v_k,\boldsymbol{F},\boldsymbol{t}}~\sum_{k=1}^{K} v_ke_k-\log v_k\label{PObj3}\\
		&~~~~\mathrm{s.t.}~~~~\boldsymbol{t}\in\boldsymbol{\Phi}_{\rm P}~\mathrm{or}~\boldsymbol{\Phi}_{\rm F}\label{Pcc3},
	\end{align}
\end{subequations}
where $e_k$ is defined as 
\begin{align}
	&e_k = \sum_{i =1,i\neq k}^{K}\big|u_k\boldsymbol{h}_{k}^{\rm H}\boldsymbol{\overline{T}}\boldsymbol{f}_{i}\big|^2   \nonumber\\
	&~~~~~~+ \big|u_k\boldsymbol{h}_{k}^{\rm H}\boldsymbol{\overline{T}}\boldsymbol{f}_{k}-1\big|^2 +\frac{|u_k|^2\sigma^2\|\boldsymbol{F}\|_{\rm F}^2}{P},
\end{align}
and $u_k$ is the receive factor of the $k$th user. 

Furthermore, we consider the antenna selection constraint in \eqref{Pcc3}. Note that \eqref{ProblemFormulation3} adapts to both the PC-REMAA and FC-REMAA by adjusting the constraint in \eqref{Pcc3}. From \eqref{ASconstaint}, the PC-REMAA includes an additional constraint in \eqref{Cons4}, which makes its design more complex compared to the FC-REMAA. Due to this added difficulty, we focus on solving \eqref{ProblemFormulation3} for the PC-REMAA in this work. Once the algorithm is developed, it can be directly applied to the FC-REMAA by simply removing the additional constraint in \eqref{Cons4}.

To handle the binary constraint in $\boldsymbol{\Phi}_{\rm P}$, we use the equivalent continuous formulation from \cite{TWC24LR}, expressed as
\begin{align}\label{eqbinary}
	[\boldsymbol{t}]_p(1 - [\boldsymbol{t}]_p) = 0,~\mbox{for}~p=1,2,\cdots,N_{\rm t}.
\end{align}
Using this, the feasible set $\boldsymbol{\Phi}_{\rm P}$ can be rewritten  as 
\begin{align}\label{RestatePhiP}
	\boldsymbol{\Phi}_{\rm P} = \{\boldsymbol{t}\in\mathbb{Z}^{N_{\rm t}}|\eqref{eqbinary},~\eqref{Cons2},~\eqref{Cons3}, \mbox{and}~\eqref{Cons4}\}.
\end{align}
Note that $\boldsymbol{\Phi}_{\rm P}$ in \eqref{RestatePhiP} contains  three equality constraints and one inequality constraint. The presence of multiple equality constraints reduces the degrees of freedom in the optimization, which may lead to lower-quality solutions. To address this issue,  we employ the penalty-based method to relax the equality constraints and incorporate them into the objective. This reformulation transforms \eqref{ProblemFormulation3} into the following problem: 
\begin{subequations}\label{ProblemFormulation4}
	\begin{align}
		&\min_{u_k,v_k,\boldsymbol{F},\boldsymbol{t}}~\mathcal{L}(u_k,v_k,\boldsymbol{F},\boldsymbol{t})\label{PObj4}\\
		&~~~~\mathrm{s.t.}~~~~\boldsymbol{S}^{\rm T}\boldsymbol{t}\le \boldsymbol{1}_{N_{\rm t}}\label{Pcc4},\\
		&~~~~~~~~~~~~[\boldsymbol{t}]_p\in[0,1],~\mathrm{for}~p=1,2,\cdots,N_{\rm t}\label{BoxConstraint}.
	\end{align}
\end{subequations}
In \eqref{PObj4}, $\mathcal{L}(u_k,v_k,\boldsymbol{F},\boldsymbol{t})$ represents a weighted sum of the original objective function and the penalty terms for the relaxed constraints, given by
\begin{align}
	\mathcal{L}(u_k,v_k,\boldsymbol{F},&\boldsymbol{t})	\triangleq \sum_{k=1}^{K} \left(v_ke_k-\log v_k\right) + \rho_1\boldsymbol{t}^{\rm T}(\boldsymbol{1}_{N_{\rm t}} - \boldsymbol{t}) \nonumber \\
	&+\rho_2\|\boldsymbol{1}_{N_{\rm t}}^{\rm T}\boldsymbol{t} - N_{\rm s}\|_2^2 + \rho_3\|\boldsymbol{Q}^{\rm T}\boldsymbol{t}- \boldsymbol{1}_{M_{\rm t}}\|_2^2,
\end{align}
where $\rho_1>0$, $\rho_2>0$, and $\rho_3>0$ are penalty coefficients that control the relaxation of the constraints. In this new formulation, the inequality constraint in \eqref{Pcc4} is retained, while an additional box constraint in \eqref{BoxConstraint} is introduced to further refine the feasible region. With this reformulation in place, we now proceed to develop the TL-JBAS algorithm to efficiently solve \eqref{ProblemFormulation4}.

\subsection{Description of the TL-JBAS Algorithm}  

The TL-JBAS algorithm follows a two-loop iterative framework to optimize beamforming and antenna selection efficiently.  

In the outer loop, the penalty coefficients \( \rho_1 \), \( \rho_2 \), and \( \rho_3 \) are initially set to 1 to provide a well-conditioned starting point. These coefficients are then gradually increased by multiplying them with scaling factors \( \beta_1 \), \( \beta_2 \), and \( \beta_3 \), each of which is greater than 1. This gradual adjustment ensures that the equality constraints are eventually satisfied while maintaining numerical stability.  

In the inner loop, the penalty coefficients remain fixed, and an alternating minimization method is employed to iteratively  optimize \( u_k \), \( v_k \), \( \boldsymbol{F} \), and \( \boldsymbol{t} \).

%
%



\subsection{Description of the Alternating Minimization Method}
\subsubsection{Initialization} First, the alternating minimization method is initialized by assuming that all antennas are selected. Accordingly, the antenna selection vector \( \boldsymbol{t} \) is set as  
\begin{align}\label{init}
	\boldsymbol{t} = \boldsymbol{1}_{N_{\rm t}}.
\end{align}
Next, to maximize the received power for each user at the start, the beamforming matrix \( \boldsymbol{F} \) is initialized based on the channel state information, given by  
\begin{align}\label{iniF}
	\boldsymbol{F} = \boldsymbol{H}.
\end{align}

\subsubsection{Optimization of \( u_k \)}\label{Optuk}  

When optimizing \( u_k \) while keeping other variables fixed, the problem in \eqref{ProblemFormulation3} simplifies to  
\begin{align}\label{OptUk}  
	\min_{u_k}~e_k.  
\end{align}  
Next, we first compute the partial derivative of \( e_k \) with respect to \( u_k^* \), given by  
\begin{align}  
	\frac{\partial e_k}{\partial u_k^*} =  \left(\sum_{i=1}^{K}\big|\boldsymbol{h}_{k}^{\rm H}\boldsymbol{\overline{T}}\boldsymbol{f}_{i}\big|^2 + \frac{\sigma^2\|\boldsymbol{F}\|_{\rm F}^2}{P}\right) - \boldsymbol{f}_{k}^{\rm H}\boldsymbol{\overline{T}}\boldsymbol{h}_{k}.  
\end{align}  
Setting the derivative to zero, i.e., \( {\partial e_k}/{\partial u_k^*} = 0 \), we obtain the optimal solution for \eqref{OptUk} as  
\begin{align}\label{Optreuk}  
	\widetilde{u}_k = \frac{\boldsymbol{f}_{k}^{\rm H}\boldsymbol{\overline{T}}\boldsymbol{h}_{k}}{\sum_{i=1}^{K}\big|\boldsymbol{h}_{k}^{\rm H}\boldsymbol{\overline{T}}\boldsymbol{f}_{i}\big|^2 + \frac{\sigma^2\|\boldsymbol{F}\|_{\rm F}^2}{P}}.  
	\end{align}

\subsubsection{Optimization of \( v_k \)}  

When optimizing \( v_k \) while keeping other variables fixed, the problem in \eqref{ProblemFormulation3} reduces to  
\begin{align}\label{OptVk}  
	\min_{v_k}~ v_k e_k - \log v_k.  
\end{align}  
To determine the optimal solution, we apply the first-order optimality condition by differentiating the objective function with respect to \( v_k \) and setting it to zero. Solving for \( v_k \), we obtain  
\begin{align}\label{Optrevk}  
	\widetilde{v}_k = \frac{1}{e_k}.  
\end{align}  

\subsubsection{Optimization of \( \boldsymbol{f}_k \)}  

When optimizing \( \boldsymbol{f}_k \) while keeping other variables fixed, the problem in \eqref{ProblemFormulation3} reduces to  
\begin{align}\label{Optfk}  
	\min_{\boldsymbol{f}_k}~ \sum_{k=1}^{K} v_k e_k.  
\end{align}  
Next, we compute the gradient of the objective function with respect to \( \boldsymbol{f}_k^* \), yielding  
\begin{align}  
	\sum_{k=1}^{K} v_k \frac{\partial e_k}{\partial \boldsymbol{f}_k^*} = \boldsymbol{\Psi} \boldsymbol{f}_k - \boldsymbol{\eta},  
\end{align}  
where the terms \( \boldsymbol{\Psi} \) and \( \boldsymbol{\eta} \) are defined as  
\begin{align}  
	&\boldsymbol{\Psi} = \sum_{i=1}^{K} v_k |u_k|^2 \left( \boldsymbol{\overline{T}}\boldsymbol{h}_k \boldsymbol{h}_k^{\rm H} \boldsymbol{\overline{T}}^{\rm T}  + \frac{\sigma^2}{P} \boldsymbol{I}_{N_{\rm t}} \right), \nonumber \\  
	&\boldsymbol{\eta} = v_k u_k\boldsymbol{\overline{T}} \boldsymbol{h}_k .  
\end{align}  
By setting \( \sum_{k=1}^{K} v_k \frac{\partial e_k}{\partial \boldsymbol{f}_k^*} = \boldsymbol{0} \), we obtain the optimal solution for \eqref{Optfk} as  
\begin{align}\label{Optref}  
	\widetilde{\boldsymbol{f}}_k = (\boldsymbol{\Psi}^{\rm H} \boldsymbol{\Psi})^{-1} \boldsymbol{\Psi}^{\rm H} \boldsymbol{\eta}.  
\end{align}  
By stacking all \( \widetilde{\boldsymbol{f}}_k \), we obtain the optimized beamforming matrix \( \widetilde{\boldsymbol{F}} \).

\subsubsection{Optimization of \( \boldsymbol{t} \)}\label{Optt}

In this part, we optimize \( \boldsymbol{t} \) while keeping other variables fixed. The objective function can be expressed as  
\begin{align}
	\mathcal{L}(u_k,v_k,\boldsymbol{F},\boldsymbol{t}) &\propto \sum_{k=1}^{K} v_k e_k 
	+ \rho_1 \boldsymbol{t}^{\rm T}(\boldsymbol{1}_{N_{\rm t}} - \boldsymbol{t}) \nonumber \\
	&~~+ \rho_2 \|\boldsymbol{1}_{N_{\rm t}}^{\rm T} \boldsymbol{t} - N_{\rm s}\|_2^2 
	+ \rho_3 \|\boldsymbol{Q}^{\rm T} \boldsymbol{t} - \boldsymbol{1}_{M_{\rm t}}\|_2^2. \nonumber 
\end{align}
where $\propto$ denotes "proportional to" indicating that terms independent of the optimization variables are omitted. Expanding \( e_k \), we rewrite the above function as  
\begin{align}
	&~~~~\mathcal{L}(u_k,v_k,\boldsymbol{F},\boldsymbol{t})\nonumber \\
	&\propto \sum_{k=1}^{K} v_k \left( \sum_{i=1}^{K} |u_k \boldsymbol{h}_k^{\rm H} \boldsymbol{\overline{T}} \boldsymbol{f}_i|^2 
	- 2\mathcal{R} \{ u_k \boldsymbol{h}_k^{\rm H} \boldsymbol{\overline{T}} \boldsymbol{f}_k \} \right) \nonumber \\
	&~~~~- \rho_1 (\boldsymbol{t}^{\rm T} \boldsymbol{t} - \boldsymbol{1}_{N_{\rm t}}^{\rm T} \boldsymbol{t})
	+ \rho_2 (\boldsymbol{t}^{\rm T} \boldsymbol{1}_{N_{\rm t}} \boldsymbol{1}_{N_{\rm t}}^{\rm T} \boldsymbol{t} 
	- 2N_{\rm s} \boldsymbol{1}_{N_{\rm t}}^{\rm T} \boldsymbol{t}) \nonumber \\
	&~~~~+ \rho_3 \boldsymbol{t}^{\rm T} \boldsymbol{Q} \boldsymbol{Q}^{\rm T} \boldsymbol{t} 
	- 2 \boldsymbol{1}_{M_{\rm t}}^{\rm T} \boldsymbol{Q}^{\rm T} \boldsymbol{t}.
\end{align}
To facilitate further analysis, we define \( \boldsymbol{w}_{k,i} \triangleq \boldsymbol{h}_k \odot \boldsymbol{f}_i \) and rewrite the objective as  
\begin{align}
	&~~~~\mathcal{L}(u_k,v_k,\boldsymbol{F},\boldsymbol{t})  \nonumber\\
	&= \sum_{k=1}^{K} v_k \left( \sum_{i=1}^{K} |u_k \boldsymbol{t}^{\rm T} \boldsymbol{w}_{k,i}|^2 
	- 2\mathcal{R} \{ u_k \boldsymbol{t}^{\rm T} \boldsymbol{w}_{k,k} \} \right) \nonumber \\
	&~~~~+ \boldsymbol{t}^{\rm T} (-\rho_1 \boldsymbol{I}_{N_{\rm T}} + \rho_2 \boldsymbol{1}_{N_{\rm t}} \boldsymbol{1}_{N_{\rm t}}^{\rm T} 
	+ \rho_3 \boldsymbol{Q} \boldsymbol{Q}^{\rm T}) \boldsymbol{t} \nonumber \\  
	&~~~~+ (\rho_1 \boldsymbol{1}_{N_{\rm t}}^{\rm T} - 2\rho_2 N_{\rm s} \boldsymbol{1}_{N_{\rm t}}^{\rm T} 
	- 2 \boldsymbol{1}_{M_{\rm t}}^{\rm T} \boldsymbol{Q}^{\rm T}) \boldsymbol{t} \nonumber \\  
	&\overset{\mathrm{(a)}}{=} \boldsymbol{t}^{\rm T} \boldsymbol{U} \boldsymbol{t} + \boldsymbol{u}^{\rm T} \boldsymbol{t},
\end{align}
where in  \( \mathrm{(a)} \), we define  
\begin{align}\label{Uu}
	&\boldsymbol{U}\triangleq \sum_{k=1}^{K}\!v_k|u_k|^2\!\sum_{i=1}^{K}\boldsymbol{w}_{k,i}^{\rm H}\boldsymbol{w}_{k,i}\!-\!\rho_1\boldsymbol{I}_{N_{\rm t}} \!+\!\rho_2\boldsymbol{1}_{N_{\rm t}}\!\boldsymbol{1}_{N_{\rm t}}\!+\!\rho_3\boldsymbol{Q}\boldsymbol{Q}^{\rm T}\nonumber\\
	&\boldsymbol{u}\triangleq -2\sum_{k=1}^{K}v_k\mathcal{R}\{u_k\boldsymbol{w}_{k,k}\} + \rho_1\boldsymbol{1}_{N_{\rm T}}-2\rho_2N_{\rm s}\boldsymbol{1}_{N_{\rm t}}-2\boldsymbol{Q}\boldsymbol{1}_{M_{\rm t}}.
\end{align}
Thus, the optimization problem reduces to  
\begin{subequations}\label{ProblemFormulation5}
	\begin{align}
		&\min_{\boldsymbol{t}}~ \boldsymbol{t}^{\rm T} \boldsymbol{U} \boldsymbol{t} + \boldsymbol{u}^{\rm T} \boldsymbol{t}, \\  
		&~\mathrm{s.t.}~~ \boldsymbol{S}^{\rm T} \boldsymbol{t} \leq \boldsymbol{1}_{N_{\rm t}}, \\  
		&~~~~~~~ [\boldsymbol{t}]_p \in [0,1],~ \forall p=1,2,\dots,N_{\rm t}.
	\end{align}
\end{subequations}
Since \( \boldsymbol{U} \) contains the term \( -\rho_1 \boldsymbol{I}_{N_{\rm T}} \) in \eqref{Uu}, it may not necessarily be positive definite.  
\begin{itemize}  
	\item If \( \boldsymbol{U} \) is positive definite, \eqref{ProblemFormulation5} becomes a convex quadratic program and can be efficiently solved using CVX.  
	\item If \( \boldsymbol{U} \) is not positive definite, \eqref{ProblemFormulation5} is nonconvex, requiring more sophisticated optimization methods such as the interior-point method. These methods are well established and can be implemented using standard optimization toolboxes. For further details, we refer the interested reader to \cite{Waltz2006}.  
\end{itemize}  
By solving \eqref{ProblemFormulation5}, we obtain the antenna selection vector \( \boldsymbol{\widetilde{t}} \).

\subsubsection{Stop Conditions}\label{SCAM} We repeat the procedures from Section \ref{Optuk} to Section \ref{Optt} to alternately optimize $u_k$, $v_k$, $\boldsymbol{F}$, and $\boldsymbol{t}$ until the maximum number of iterations is reached or the alternating minimization method converges.

\subsection{Stopping Conditions of the TL-JBAS Algorithm}\label{SCTL}

To determine whether an antenna is selected, we introduce a threshold \( \varUpsilon \). Specifically, if \( [\boldsymbol{\widetilde{t}}]_n \geq \varUpsilon \), the \( n \)th antenna is selected; otherwise, it is not included in the final selection. 

We iteratively update the penalty parameters \( \rho_1 \), \( \rho_2 \), and \( \rho_3 \) in the outer loop and perform the alternating minimization method in the inner loop, until the number of selected antennas equals \( N_{\rm s} \). The final selection vector is
\begin{align}\label{widehatt}
	[\widehat{\boldsymbol{t}}]_n = 
	\begin{cases}
		1, & [\widetilde{\boldsymbol{t}}]_n \geq \varUpsilon,	\\
		0, & \text{otherwise}.
	\end{cases}
\end{align}
Substituting \( \widehat{\boldsymbol{t}} \) into the original problem in \eqref{ProblemFormulation}, we reformulate the optimization problem as
\begin{align}\label{ProblemFormulation6}
	&\max_{\boldsymbol{F}}~\sum_{k=1}^{K}  \log_2\left(1+\frac{\big|\boldsymbol{h}_{k}^{\rm H}\mathrm{diag}(\widehat{\boldsymbol{t}})\boldsymbol{f}_{k}\big|^2}{\sigma^2+\sum_{i =1,i\neq k}^{K}|\boldsymbol{h}_{k}^{\rm H}\mathrm{diag}(\widehat{\boldsymbol{t}})\boldsymbol{f}_{i}\big|^2} \right) \nonumber\\
	&~\mathrm{s.t.}~~\left\|\boldsymbol{F}\right\|_{\rm F}^2 \le P.
\end{align}
The optimization problem in \eqref{ProblemFormulation6} is a multiuser sum-rate maximization problem, which can be effectively solved using the weighted minimum mean squared error (WMMSE) framework~\cite{TSP23ZXT,TWC15SQJ}. As WMMSE-based solutions are well-established, we omit the detailed derivation and denote the resulting maximized sum-rate as \( V \).

\begin{algorithm}[!t]
	\caption{Two-Step Multiuser Beamforming and Antenna Selection (TS-MBAS) Scheme}
	\label{alg_RHB}
	\begin{algorithmic}[1]
		\STATE \textbf{Input:} $N_{\rm r}$, $N_{\rm c}$, $N_{\rm t}$, $S_{\rm r}$, $S_{\rm c}$, $Q_{\rm r}$, $Q_{\rm c}$, $P$, $K$, $\varUpsilon$, and  $\boldsymbol{h}_k$.
		\STATE \textbf{Initialization:} $\rho_1\leftarrow 1$,~ $\rho_2\leftarrow 1$,~\mbox{and}~ $\rho_3\leftarrow 1$.
		\STATE /*\textit{Step 1: The TL-JBAS Algorithm}*/
		\WHILE{conditions in Sec.~\ref{SCTL} are not satisfied}
		\STATE $\rho_1\leftarrow \rho_1\beta_1$,~ $\rho_2\leftarrow \rho_2\beta_2$,~\mbox{and}~ $\rho_3\leftarrow \rho_3\beta_3$.
		\STATE /*\textit{The Alternating Minimization Method}*/
		\WHILE{conditions in Sec.~\ref{SCAM} are not satisfied}
		\STATE Initialize  $\boldsymbol{t}$ and $\boldsymbol{F}$ via \eqref{init} and \eqref{iniF}, respectively.
		\STATE Obtain $\widetilde{u}_k$ via \eqref{Optreuk}.
		\STATE Obtain  $\widetilde{v}_k$ via \eqref{Optrevk}.
		\STATE Obtain  $\widetilde{\boldsymbol{f}}_k$ via \eqref{Optref}.
		\STATE Obtain  $\widetilde{\boldsymbol{t}}$ via \eqref{ProblemFormulation5}.
		\ENDWHILE
		\ENDWHILE
		\STATE Obtain $\widehat{\boldsymbol{t}}$ via \eqref{widehatt}.
		\STATE /*\textit{Step 2:  Enhancement}*/
		\STATE $\overline{\boldsymbol{t}}_0 \leftarrow \widehat{\boldsymbol{t}}$ and ~$d\leftarrow 0$.
		\WHILE{conditions in \eqref{StopConditions} are not satisfied}
		\STATE $d\leftarrow d + 1$.
		\STATE Obtain $n$ via \eqref{IndexAntenna}.
		\STATE Obtain $\overline{p}_{d}^{(n)}$ via \eqref{PbN}.
		\STATE Obtain $\widehat{V}_d$ by solving \eqref{ProblemFormulation6}.
		\ENDWHILE
		\STATE $\widetilde{V}\leftarrow \widehat{V}_d$.
		\STATE \textbf{Output:} $\widetilde{V}$.
	\end{algorithmic}
\end{algorithm}

\subsection{Enhancement with the Coordinate Descent Method}\label{enhancement}

In this part, we present the second step of the TS-MBAS scheme, where the coordinate descent method is employed to enhance the performance of the TL-JBAS algorithm. The core idea is to iteratively evaluate the selected antennas by sequentially swapping each selected antenna with a candidate antenna.  Specifically, when each selected antenna is replaced by a candidate antenna, a multiuser sum-rate maximization problem is solved and the maximized sum-rate is calculated. If the replacement improves the sum-rate, the swap is accepted; otherwise, the original selection is retained. This process continues until no further sum-rate improvement can be achieved through any single swap.

According to Section~\ref{SCTL}, the TL-JBAS algorithm selects $N_{\rm s}$ antennas. We denote the index of the $n$th selected antenna as $I_n$, for $n = 1,2,\dots,N_{\rm s}$. Then we denote the initial antenna selection vector as $\overline{\boldsymbol{t}}_0 \leftarrow \widehat{\boldsymbol{t}}$ and  iteratively evaluate the selected antennas.

\subsubsection{Iterative Antenna Swapping}
In the $d$th iteration, for $d \geq 1$, we initialize the antenna selection vector as $\overline{\boldsymbol{t}}_{d} \leftarrow \overline{\boldsymbol{t}}_{d-1}$. The index of the antenna under evaluation is given by
\begin{align}\label{IndexAntenna}
	n = \mathrm{mod}(d-1, N_{\rm s}) + 1.
\end{align}
Next, we deactivate the $I_n$th antenna by setting
\begin{align}
	[\overline{\boldsymbol{t}}_{d}]_{I_n} = 0.
\end{align}
We then sequentially evaluate all $N_{\rm t}$ candidate antennas. When evaluating the $p$th candidate antenna, for $p =1 ,2,\cdots,N_{\rm t}$, the updated selection vector is given by
\begin{align}
	[\overline{\boldsymbol{t}}_d]_p \leftarrow 1.
\end{align}

\subsubsection{Sum-Rate Evaluation}
For each selection vector $\overline{\boldsymbol{t}}_d$, we verify whether it belongs to the feasible set $\boldsymbol{\Phi}_{\rm P}$ or $\boldsymbol{\Phi}_{\rm F}$. If so, we substitute $\widehat{\boldsymbol{t}}$ in \eqref{ProblemFormulation6} with $\overline{\boldsymbol{t}}_d$ and solve it using the WMMSE approach. The resulting sum-rate for this evaluation is denoted as $\overline{V}_p$. If the selection is infeasible, we set $\overline{V}_p \leftarrow 0$. 

The index of the antenna yielding the maximum sum-rate for the evaluation of the $n$th antenna in the $d$th iteration is determined as
\begin{align}\label{PbN}
	\overline{p}_{d}^{(n)} = \arg\max_{p=1,2,\dots,N_{\rm t}}~\overline{V}_p.
\end{align}
The corresponding sum-rate for the $d$th iteration is denoted as $\widehat{V}_d$. The index  of the $n$th selected antennas, $I_n$, is then updated as
\begin{align}\label{IndexAntennaUpdate}
	I_n \leftarrow \overline{p}_{d}^{(n)}.
\end{align}

\subsubsection{Stopping Criterion}
The iterative process continues following the steps from \eqref{IndexAntenna} to \eqref{IndexAntennaUpdate} until the stopping condition
\begin{align}\label{StopConditions}
	\overline{p}_{d}^{(n)} = \overline{p}_{d-N_{\rm s}}^{(n)}, \quad  d > N_{\rm s}, \quad n = 1,2,\dots,N_{\rm s},
\end{align}
is satisfied. This condition implies that the selected antennas remain unchanged compared to those chosen $N_{\rm s}$ iterations earlier, indicating that no further improvement can be achieved through any single swap.

\subsubsection{Final Optimization Outcome}
Upon convergence, the optimized sum-rate after enhancement is obtained as
\begin{align}
	\widetilde{V} \leftarrow \widehat{V}_d,
\end{align}
which completes the enhancement procedures.

Finally, we summarize the  TS-MBAS scheme, including the TL-JBAS algorithm in the first step and the enhancement using the coordinate descent method in the second step,  in \textbf{Algorithm~\ref{alg_RHB}}.

\section{The Mechanical Movable Antennas: 	A Revisit}\label{MMA}

In this section, to establish a benchmark for evaluating REMA-enabled communications, we revisit MMAs with continuously adjustable positions within the transmission region. We formulate a sum-rate maximization problem for MMA-enabled multiuser communications and propose an ABAPO scheme to solve it.

\subsection{System and Channel Model}
For a fair comparison with REMAs,  we assume that the number of antennas in the MMA system is $N_{\rm s}$. The three-dimensional coordinate of the $n$th antenna is denoted as $\boldsymbol{p}_n \in\mathbb{R}^3$, for $n=1,2,\dots,N_{\rm s}$. The transmission region of the MMAs is identical to that of the REMAs and is denoted as $\mathcal{C}$, indicating that the antenna positions satisfy the constraint $\boldsymbol{p}_n \in \mathcal{C}$. 

From \eqref{channel}, the channel between the $k$th user and the $N_{\rm s}$ antennas can be expressed as 
\begin{align}
	\boldsymbol{g}_{k}(\boldsymbol{P}) =  \sum_{l=1}^{L_{k}}\gamma_{k}^{(l)}\boldsymbol{\beta}\big(N_{\rm s},\boldsymbol{P},\theta_k^{(l)},\phi_k^{(l)}\big).
\end{align}
$\boldsymbol{\beta}\big(N_{\rm s},\theta,\phi\big)$ denotes the channel steering vector between the $k$th user and the $N_{\rm s}$ movable antennas, and is given by
\begin{align}
	\boldsymbol{\beta}\big(N_{\rm s},\boldsymbol{P},\theta,\phi\big) = \sqrt{\frac{1}{N_{\rm s}}} e^{j2\pi \boldsymbol{P}\boldsymbol{i}},
\end{align}
where  $\boldsymbol{P} \triangleq [\boldsymbol{p}_1,\boldsymbol{p}_2,\dots,\boldsymbol{p}_{N_{\rm s}}]^{\rm T}$ is the stacked position matrix for the $N_{\rm s}$ MMAs and  $\boldsymbol{i} \triangleq [\cos\phi\sin\theta,\cos\phi\cos\theta,\sin\phi]^{\rm T}$ is the normalized wave vector.

\subsection{Problem Formulation}

Each of the $N_{\rm s}$ antennas is connected to an RF chain. By designing the baseband beamforming vector $\boldsymbol{w}_k \in\mathbb{C}^{N_{\rm s}}$ for each user, we can mitigate multiuser interference and optimize the multiuser sum-rate. Additionally, by adjusting the positions of the $N_{\rm s}$ MMAs via motors or liquid metals, the multiuser channels can be tuned to achieve appropriate coherence, thus facilitating effective beamforming.

We aim at maximizing the multiuser sum-rate by jointly optimizing the positions of the MMAs, $\boldsymbol{p}_n$, and the baseband beamforming vectors, $\boldsymbol{w}_k$. The optimization problem is formulated as
\begin{subequations}\label{MultiuserSumRate}
	\begin{align}
		&\max_{\boldsymbol{p}_n,\boldsymbol{w}_k}~\sum_{k=1}^{K} \overline{R}_k\label{Obj2}\\
		&~\mathrm{s.t.}~~~\left\|\boldsymbol{W}\right\|_{\rm F}^2 \le P,\label{Con21}\\
		&~~~~~~~~~\boldsymbol{p}_n\in\mathcal{C},~n=1,2,\dots,N_{\rm s},\label{Con22}\\
		&~~~~~~~~~\|\boldsymbol{p}_m-\boldsymbol{p}_n\|_2\ge\frac{\lambda}{2},~\forall m\neq n.\label{Con23}
	\end{align}
\end{subequations}
In \eqref{Obj2}, $\overline{R}_k$ represents the achievable sum-rate for the $k$th user and can be expressed as
\begin{align}
	\overline{R}_k = \log_2\left(1+\frac{\big|\boldsymbol{g}_{k}(\boldsymbol{P})^{\rm H}\boldsymbol{w}_{k}\big|^2}{\sigma^2+\sum_{i =1,i\neq k}^{K}|\boldsymbol{g}_{k}(\boldsymbol{P})^{\rm H}\boldsymbol{w}_{i}\big|^2} \right).
\end{align}
In \eqref{Con21}, $\boldsymbol{W} \triangleq [\boldsymbol{w}_1,\dots,\boldsymbol{w}_{K}]$ is the stack of beamforming vectors, and its Frobenius norm is constrained by the total transmit power $P$.
In \eqref{Con22}, the position of each antenna $\boldsymbol{p}_n$ is restricted to lie within the predefined transmission region $\mathcal{C}$.
In \eqref{Con23}, the spacing between any two antennas is at least $\lambda/2$ to avoid mutual coupling effects.

The optimization problem in \eqref{MultiuserSumRate} involves the optimization of antenna positions $\boldsymbol{p}_n$ and user beamforming vectors $\boldsymbol{w}_k$. These two sets of variables exhibit different characteristics and require distinct optimization techniques. Therefore, it is difficult to optimize them simultaneously. To address this, we then propose an ABAPO scheme to  efficiently solve \eqref{MultiuserSumRate} by alternately optimizing  the multiuser beamforming and the antenna positions.

\subsection{ Alternating Beamforming and Antenna Position Optimization Scheme}
Given the antenna positions $\boldsymbol{p}_n$, the optimization problem in \eqref{MultiuserSumRate} simplifies  to
\begin{align}\label{MultiuserSumRate2}
	&\max_{\boldsymbol{w}_k}~\sum_{k=1}^{K} \overline{R}_k\nonumber\\
	&~\mathrm{s.t.}~~~\left\|\boldsymbol{W}\right\|_{\rm F}^2 \le P.
\end{align}
This is a classic multiuser sum-rate maximization problem and can be efficiently solved using the WMMSE framework. We omit the details and denote its solution as $\boldsymbol{\widetilde{w}}_k$.

Given the  beamforming vectors $\boldsymbol{w}_k$, \eqref{MultiuserSumRate} reduces to
\begin{align}\label{AntennaPositionOpt}
	&\max_{\boldsymbol{p}_n}~\sum_{k=1}^{K} \overline{R}_k\nonumber\\
	&~\mathrm{s.t.}~~~\boldsymbol{p}_n\in\mathcal{C},~n=1,2,\dots,N_{\rm s},\nonumber\\
	&~~~~~~~~\|\boldsymbol{p}_m-\boldsymbol{p}_n\|_2\ge\frac{\lambda}{2},~\forall m\neq n.
\end{align}
This problem is nonconvex due to the structures of both its objective function and constraints~\cite{TWC24MWY}.  Existing works usually use the successive convex approximation method to address this nonconvex optimization. Specifically, the objective is addressed with  Taylor’s theorem to successively obtain quadratic surrogate functions. For the quadratic surrogate function, the nonconvex constraint is dealt with the Cauchy-Schwartz inequality to successively obtain linear constraints. Then, the convex optimization is used to solve the convex problem with quadratic objective function and linear constraints. However, this approach needs to solve a large amount of convex optimization problems and involves high computational complexity.


To avoid the high computational complexity of successive convex approximations, we adopt an interior-point method to solve \eqref{AntennaPositionOpt}~\cite{Waltz2006}. First, we rewrite the transmit region constraint $\boldsymbol{p}_n\in\mathcal{C}$ as $E(\boldsymbol{p}_n)\le0$ by parameterizing $\mathcal{C}$. Introducing slack variables $c_n\ge0$ and $g_{m,n}\ge0$, we reformulate \eqref{AntennaPositionOpt} as:
\begin{align}\label{AntennaPositionOpt2}
	&\min_{\boldsymbol{p}_n}~-\sum_{k=1}^{K}\overline{R}_k -\mu\sum_{n=1}^{N_{\rm s}}\ln c_n -\rho\sum_{m=1}^{N_{\rm s}}\sum_{n=m+1}^{N_{\rm s}}\ln g_{m,n} \nonumber\\
	&~\mathrm{s.t.}~~~E(\boldsymbol{p}_n) + c_n = 0,~n=1,2,\dots,N_{\rm s},\nonumber\\
	&~~~~~~~~\frac{\lambda}{2} - \|\boldsymbol{p}_m-\boldsymbol{p}_n\|_2 + g_{m,n} = 0.
\end{align}
The optimization in \eqref{AntennaPositionOpt2} is then solved using Newton’s method with backtracking line search to iteratively update $\boldsymbol{p}_n$, $c_n$, and $g_{m,n}$ until the Karush-Kuhn-Tucker (KKT) conditions are satisfied. This procedure is well-established and can be implemented using commercial software such as MATLAB. We omit the details and denote the solution as $\widetilde{\boldsymbol{p}}_{n}$.

We iteratively solve \eqref{MultiuserSumRate2} and \eqref{AntennaPositionOpt2} until the maximum number of iterations is reached or the ABAPO scheme converges. The final solutions to the nonconvex optimization in \eqref{MultiuserSumRate} are denoted as $\widehat{\boldsymbol{w}}_{k}$ and $\widehat{\boldsymbol{p}}_{n}$.

\section{Performance Gap Analysis Between REMAs and MMAs}\label{Analysis2}

In MMAs, antennas can move continuously within the transmission region, whereas in REMAs, the antenna positions are restricted to discrete locations. As a result, a performance gap arises between REMAs and MMAs, primarily caused by the interval between the candidate radiation positions. This observation leads to a fundamental question: how fine should the position interval be for REMAs to  approximate the performance of MMAs? To address this, in this section,  we derive the maximum  power loss of REMAs compared to MMAs under any given position intervals.

\begin{table*}[!t]
	\centering
	\caption{Evaluation of the Effectiveness of the Quantization Analysis for the REMAs}
	\renewcommand{\arraystretch}{1.5} 
	\begin{tabular}{ccccccccccc} 
		\toprule
		Position Interval & $0.5\lambda$ & $0.45\lambda$ & $0.40\lambda$ & $0.35\lambda$ & $0.30\lambda$ & $0.25\lambda$ & $0.20\lambda$ & $0.15\lambda$ & $0.10\lambda$ & $0.05\lambda$ \\ 
		\midrule
		Maximum Power Loss from \eqref{Powerloss2}& 59.47\%  & 51.19\%  & 42.72\%  & 34.34\%  & 26.32\%  & 18.94\%  & 12.49\%  & 7.19\%  & 3.25\% & 0.82\% \\ 
		Percentage of Cases Satisfying \eqref{Powerloss2}& 99.96\%  & 99.85\%  & 99.61\%  & 98.90\%  & 98.51\%  & 98.14\%  & 97.43\%  & 97.07\%  & 96.41\% & 96.17\% \\ 
		\bottomrule
	\end{tabular}
	\label{three_line_full_bar}
\end{table*}

\subsection{Analysis on Power Loss of  Electronic Movable Antennas}\label{Analysis}
Suppose the $k$th user transmits a normalized signal to the BS through the channel given  in \eqref{channel}. The received signal at the location $(x_n,0,z_m)$ is given by 
\begin{align}\label{receivedsignal}
	y  & = \sum_{l=1}^{L_{k}} \sqrt{\frac{1}{N_{\rm t}}} \gamma_k^{(l)} e^{j2\pi(x_n\cos\phi_k^{(l)}\sin\theta_k^{(l)} + z_m\sin\phi_k^{(l)})/\lambda} \nonumber\\
	& \overset{(\mathrm{a})}{=} \sum_{l=1}^{L_{k}} \sqrt{\frac{1}{N_{\rm t}}} \gamma_k^{(l)} e^{j2\pi\big(\Theta_k^{(l)} x_n + \Omega_k^{(l)} z_m\big)/\lambda},
\end{align}
where we define $\Theta_k^{(l)} \triangleq \cos\phi_k^{(l)}\sin\theta_k^{(l)}$ and $\Omega_k^{(l)} \triangleq \sin\phi_k^{(l)}$ in $(\mathrm{a})$. By  basic trigonometric properties, we have $\Theta_k^{(l)}\in[-1,1]$ and $\Omega_k^{(l)}\in [-1,1]$.  From \eqref{receivedsignal}, we find that the received signal depends on the positions along both the $x$-axis and $z$-axis. However, the influence of position changes along these two axes is fundamentally similar. Therefore, we fix the position along the $z$-axis and focus on variations along the $x$-axis.  Then, we rewrite \eqref{receivedsignal} as 
\begin{align}\label{receivedsignal21}
	y = \sum_{l=1}^{L_{k}} \widetilde{\gamma}_k^{(l)} e^{j2\pi\Theta_k^{(l)}x_n/\lambda},
\end{align}
where the effective channel coefficient $\widetilde{\gamma}_k^{(l)}$ is defined as $\widetilde{\gamma}_k^{(l)} \triangleq \sqrt{\frac{1}{N_{\rm t}}} \gamma_k^{(l)} e^{j2\pi\Omega_k^{(l)}z_m/\lambda}$. Note that \eqref{receivedsignal21} is structurally similar to the DTFT but with nonuniform time-domain indices. To convert \eqref{receivedsignal21} into the standard DTFT form, we uniformly quantize the angular space $[-1,1]$ into $F$ discrete samples with the $f$th sample given by $1 - 2f/F$.
Each channel angle $\Theta_k^{(l)}$ is then assigned to its nearest quantized angle, with the corresponding index denoted as $\overline{f}_l$. Due to the quantization, $\Theta_k^{(l)}$ is not exactly equal to $1 - 2\overline{f}_l/F$. However, as $F$ increases, the quantization error decreases, and $\Theta_k^{(l)}$ converges to $1 - 2\overline{f}_l/F$.

Next, we define an all-zero vector $\boldsymbol{b} \in \mathbb{C}^{F}$ and assign values to it as $[\boldsymbol{b}]_{\overline{f}_l} \leftarrow \widetilde{\gamma}_k^{(l)}$. With this definition, \eqref{receivedsignal21} can be rewritten as
\begin{align}\label{receivedsignal3}
	y  &= \sum_{f=1}^{F} [\boldsymbol{b}]_f e^{j2\pi(1 - 2f/F)x_n/\lambda}\nonumber\\
	&= e^{j2\pi x_n/\lambda} \sum_{f=1}^{F} [\boldsymbol{b}]_f e^{-j\pi f\widetilde{x}_n},
\end{align}
where we define $\widetilde{x}_n \triangleq \frac{4x_n}{\lambda F}$.
From \eqref{receivedsignal3}, it is evident that $y$ is the DTFT of $\boldsymbol{b}$, where $\widetilde{x}_n$ represents the frequency domain. According to Fourier analysis~\cite{Folland1992}, the minimum bandwidth of an $F$-point signal can be approximated using the Dirichlet kernel
\begin{align}\label{Dirichletkernel}
	G_F(\omega) = \left| \frac{\sin(F\omega/2)}{\sin(\omega/2)} \right|.
\end{align}
By numerically evaluating \eqref{Dirichletkernel}, the $\kappa$-dB bandwidth can be determined as $\Gamma_{F}^{(\kappa)}$. For example, the minimum 3-dB bandwidth of an $F$-point signal is approximately $1.76/F$. 

On the other hand, since $\boldsymbol{b}$ contains zeros at unassigned positions, its effective length can be computed as
\begin{align}
	\overline{F} = \left\lceil \frac{(\overline{\Theta}_{\rm max} - \overline{\Theta}_{\rm min})F}{2} \right\rceil,
\end{align}
where $\overline{\Theta}_{\rm max} \triangleq \max_{l} \Theta_k^{(l)}$ and $\overline{\Theta}_{\rm min} \triangleq \min_{l} \Theta_k^{(l)}$. Additionally, from \eqref{Dirichletkernel}, we observe that $x_n$ is scaled by a factor of $F\lambda/4$ relative to $\widetilde{x}_n$. Therefore, the $\kappa$-dB mainlobe width of $y$ with respect to $x_n$ is given by
\begin{align}
	B^{(\kappa)} = \Gamma_{\overline{F}}^{(\kappa)} {F\lambda}/{4}.
\end{align}
Since $\Gamma_{F}^{(\kappa)}$ decreases as $F$ increases, a larger interval between the maximum and minimum channel angles results in a smaller mainlobe width of $y$.

Furthermore, $B^{(\kappa)} \geq \Gamma_{F}^{(\kappa)} {F\lambda}/{4}$ due to $\overline{F} \leq F$. Thus, the maximum $\kappa$-dB mainlobe width of $y$ with respect to $x_n$ is
\begin{align}\label{Powerloss2}
	\overline{B}^{(\kappa)} = \Gamma_{F}^{(\kappa)} {F\lambda}/{4}.
\end{align}
Note that \eqref{Powerloss2} establishes a relationship between power loss and position interval. For instance, setting $\kappa = 3$, corresponding to a half-power loss, the position interval is $0.44\lambda$. Conversely, if the position interval is set to $0.2\lambda$, the maximum power loss is computed as  $12.5\%$. In other words, to limit power loss to at most $\kappa$ dB, the position interval should not exceed $\Gamma_{F}^{(\kappa)} F\lambda/4$.

\subsection{Evaluation of the Analysis}
%
%

Now, we  validate  the analysis in Section~\ref{Analysis} through simulations. We fix the $z$-axis and change positions along the $x$-axis within  $[-5\lambda,5\lambda]$. The position interval ranges from $0.5\lambda$ to $0.05\lambda$ in a step of $0.05\lambda$, resulting in the number of candidate antenna positions varying from 10 to 100. The number of channel paths is set to $L_k = 20$, with channel gains following $\gamma_{k}^{(l)}\sim\mathcal{CN}(0,1)$ and channel angles distributed as $\Theta_{k}^{(l)}\in[-1,1]$.  

Using \eqref{Powerloss2}, we determine the position corresponding to the maximum received power for the REMAs. Then, we refine this position using gradient descent to obtain the optimal positions for the MMAs and compute their corresponding maximum received power. Subsequently, the power loss of the REMAs compared to the MMAs can be computed directly.

In Table~\ref{three_line_full_bar}, we first calculate the maximum power loss of the REMAs compared to the MMAs using \eqref{Powerloss2}. Next, we conduct $10^5$ Monte Carlo simulations to obtain the actual power loss of the REMAs relative to the MMAs. Additionally, we compute the ratio of cases where the actual power loss remains below the theoretical maximum given by \eqref{Powerloss2}. The results in the table demonstrate that the maximum power loss predicted by our analysis holds in the vast majority of cases, which verifies the accuracy and effectiveness of our analysis for REMAs.

\section{Simulation Results}\label{SR}

Now, we  evaluate the performance of the proposed FC-REMAA and PC-REMAA, using FPAs and MMAs as benchmarks. The width and height of each REMA in the PC-REMAA are set to $\lambda$. For a fair comparison, FPAs and MMAs employ the same number of antennas as the selected antennas in the FC-REMAA and PC-REMAA. In addition, the antenna spacing in FPAs is $\lambda/2$, and the transmit region of the MMAs is the same as the antenna panels of the FC-REMAA and PC-REMAA.



\begin{figure}[t]
	\centering
	\includegraphics[width=78mm]{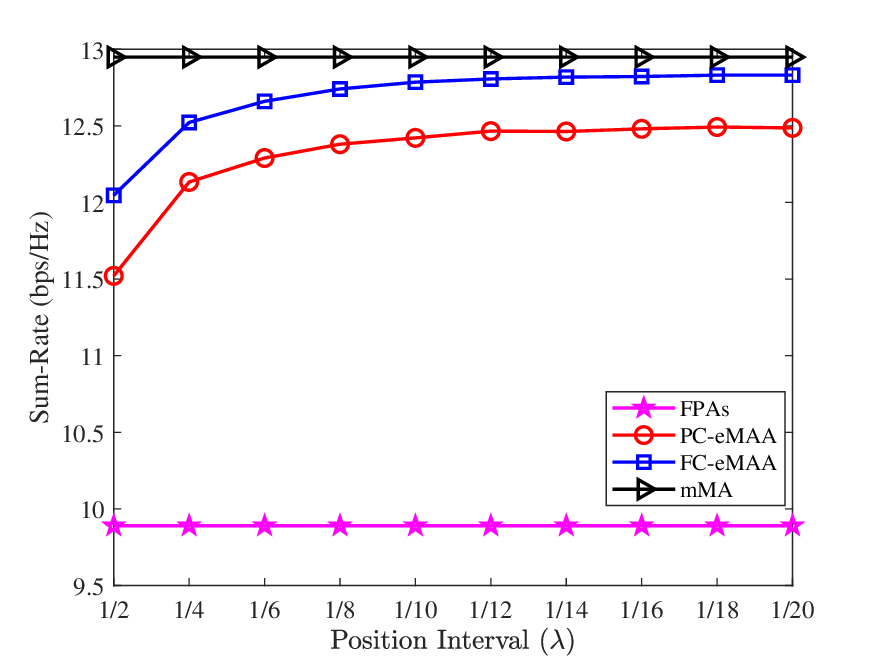}
	\caption{Comparisons of different methods in terms of the sum-rate for varying position intervals under the uniform linear array configuration. }
	\label{Fig6}
\end{figure}

\begin{figure}[t]
	\centering
	\includegraphics[width=78mm]{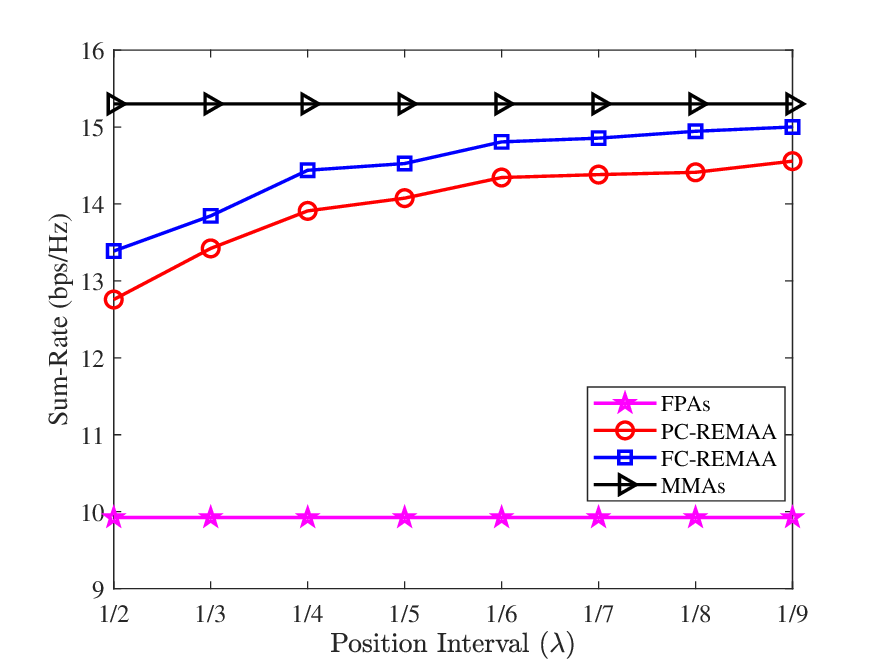}
	\caption{Comparisons of different methods in terms of the sum-rate for varying position intervals under the uniform planar array configuration. }
	\label{Fig7}
\end{figure}

\begin{figure}[t]
	\centering
	\includegraphics[width=78mm]{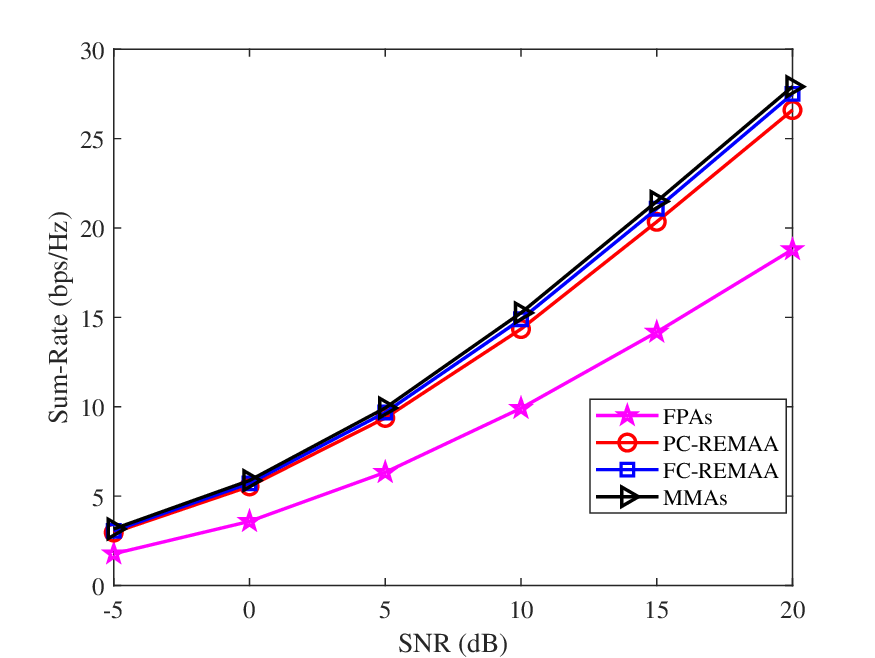}
	\caption{Comparisons of different methods in terms of the sum-rate for varying SNRs under the uniform planar array configuration. }
	\label{Fig8}
\end{figure}

Fig.~\ref{Fig6} illustrates the sum-rate performance of different methods for various position intervals in a uniform linear array configuration, where $N_{\rm r} = 1$. The number of candidate antennas in each REMA ranges from 2 to 20, corresponding to the position interval from $\lambda/2$ to $\lambda/20$. The number of selected antennas is $N_{\rm s} = 4$, the SNR is set to 10 dB, and the system includes $K = 4$ users, each with $L_{k} = 6$ channel paths. From the figure, we observe that MMAs achieve the highest sum-rate, followed by FC-REMAA, PC-REMAA, and FPAs. This performance ordering aligns with the degree of design flexibility in these arrays. Additionally, as the position interval decreases, the FC-REMAA gradually approaches the performance of MMAs. However, reducing the position interval beyond $\lambda/8$ results in only marginal improvements, which indicates that refining the position interval beyond $\lambda/8$ offers limited practical benefit in enhancing multiuser sum-rate.

\begin{figure}[!t]
	\centering
	\includegraphics[width=78mm]{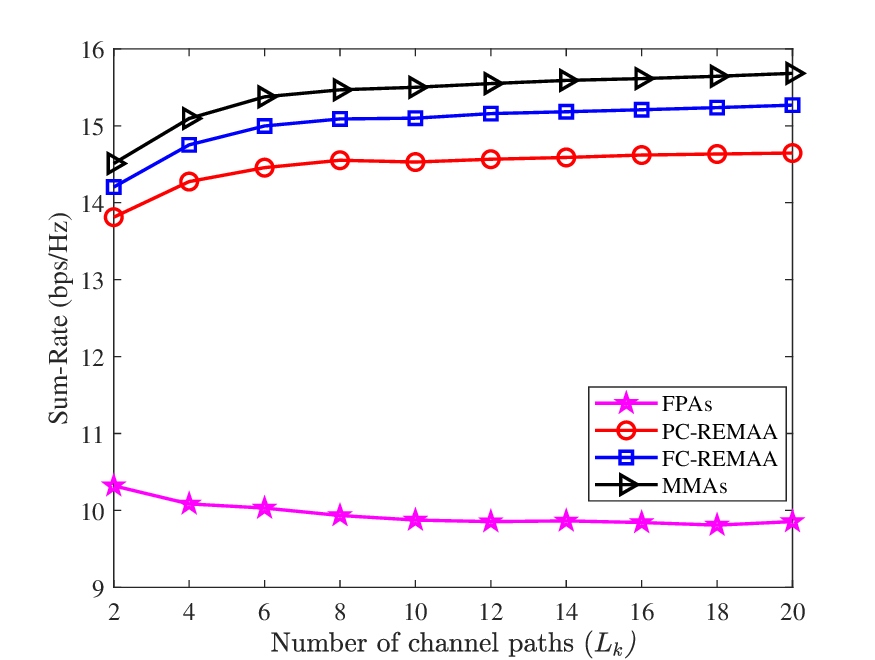}
	\caption{Comparisons of different methods in terms of the sum-rate for varying numbers of channel paths under the uniform planar array configuration.  }
	\label{Fig9}
\end{figure}

\begin{figure}[!t]
	\centering
	\includegraphics[width=78mm]{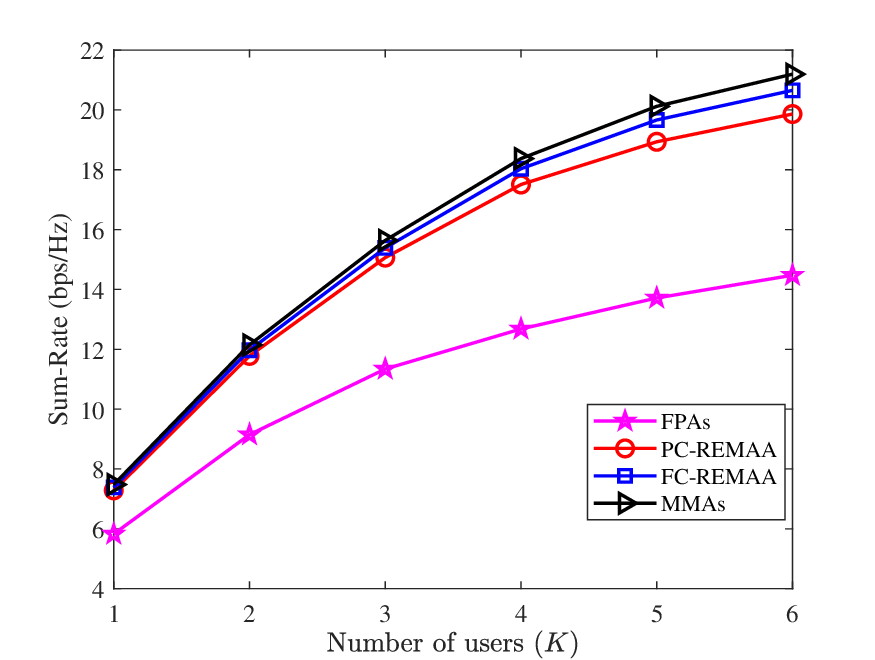}
	\caption{Comparisons of different methods in terms of the sum-rate for varying numbers of users under the uniform planar array configuration. }
	\label{Fig10}
\end{figure}


Fig.~\ref{Fig7} extends the analysis in Figure~\ref{Fig6} to a uniform planar array configuration. The number of candidate antennas per row and column in each REMA ranges from 2 to 9, corresponding to the position interval from $\lambda/2$ to $\lambda/9$. The system parameters remain consistent with those in Figure~\ref{Fig6}. Comparing the results from Figures~\ref{Fig6} and~\ref{Fig7}, we find that the MMAs, FC-REMAA, and PC-REMAA perform better in the planar configuration due to the increased flexibility in antenna placement. In contrast, the performance of FPAs remains similar in both figures because the fixed antenna positions limit their adaptability. Furthermore, as the position interval decreases, the FC-REMAA can also gradually approach the performance of MMAs in uniform planar array configurations.



Fig.~\ref{Fig8} illustrates  the sum-rate performance under varying SNRs, considering a uniform planar array configuration with eight candidate antennas per row and column in each REMA. The number of selected antennas is $N_{\rm s} = 4$, and the system includes $K = 4$ users, each with $L_{k} = 6$ channel paths. The results show that the performance gap between MMAs and FC-REMAA widens gradually as the SNR increases, while the gap between FC-REMAA and FPAs grows more rapidly.  This trend is due to the significantly higher flexibility of FC-REMAA compared to FPAs in antenna position design, which allows for more effective suppression of multiuser interference. As the SNR increases and noise power diminishes, interference becomes the dominant factor affecting performance, making interference mitigation even more critical. Consequently, the advantage of FC-REMAA over FPAs becomes more apparent. However, while MMAs offer even greater flexibility than FC-REMAA, the additional  freedom is comparatively  limited, leading to a slower performance gain relative to FC-REMAA as the SNR increases. Moreover, FC-REMAA consistently approaches the performance of MMAs across different SNR levels, demonstrating the effectiveness of the proposed approach.

Fig.~\ref{Fig9} illustrates  the impact of the number of channel paths on sum-rate performance in the uniform planar array configuration. The number of candidate antennas per row and column in an REMA is set to 8,  the number of selected antennas is $N_{\rm s} = 4$, and the SNR remains at 10~dB. From the figure, the performance of MMAs, FC-REMAA, and PC-REMAA improves as the number of channel paths increases because the received power fluctuates more rapidly within the receive region. In this context, these methods can more easily identify favorable antenna positions for multiuser communications. However, the rapid power variation also results in a narrower mainlobe width and greater power loss for FPAs, as analyzed in Section~\ref{Analysis2}, which 
leads to the decline in performance of FPAs as the number of channel paths increases.



Fig.~\ref{Fig10} illustrates  the sum-rate performance for varying  number of users under a uniform planar array configuration. Each REMA consists of 6 candidate antennas per row and column. The PC-REMAA employs three REMAs per row and two REMAs per column, leading to a total of 18 candidate antennas per row and 12 candidate antennas per column. For fairness, FC-REMAA is configured with the same number of candidate antennas as the PC-REMAA. The number of selected antennas is $N_{\rm s} = 6$, the SNR is set to 10 dB, and each user has $L_{k} = 6$ channel paths. The results show that as the number of users increases, the performance gap between MMAs and FC-REMAA widens gradually, whereas the gap between FC-REMAA and FPAs expands more rapidly. This is because greater antenna position flexibility enhances sum-rate performance, particularly in complicated scenarios with a larger number of users.

Across Fig.~\ref{Fig6} to Fig.~\ref{Fig10}, a consistent trend emerges: MMAs achieve the best performance, followed by FC-REMAA, PC-REMAA, and FPAs. This ranking aligns with the level of design flexibility available in each approach. Although FC-REMAA does not achieve exactly the same performance as MMAs, its performance remains close under various conditions, demonstrating its potential as a practical alternative with significantly reduced complexity.

\section{Conclusion}\label{CC}
In this paper, we have investigated  REMAs for multiuser communications. We have modeled each REMA as an antenna characterized by a set of predefined and discrete selectable radiation positions within the radiating region. Considering trade-off between performance and cost, we have proposed two types of REMA-based arrays: the PC-REMAA and FC-REMAA. We have formulated a multiuser sum-rate maximization problem subject to power constraint and hardware constraints of the PC-REMAA or FC-REMAA. To solve this problem, we have proposed a TS-MBAS scheme. In addition, we have revisited  MMAs  with continuously adjustable positions within the transmission region to establish a benchmark for evaluating REMA-enabled multiuser communications. We have analyzed the performance gap between REMAs and MMAs. Specifically, we have  transformed the received signal of MMAs into the DTFT of the channel coefficients. Based on Fourier analysis, we have derived the maximum power loss of REMAs compared to MMAs for any given position interval. Future research will focus on further exploring REMA applications in wireless communications.


\bibliographystyle{IEEEtran}
\bibliography{IEEEabrv,IEEEexample}

\end{document}